\documentclass[aps, reprint, pra, nofootinbib]{revtex4-2}

\usepackage{hyperref}

\makeatletter 
    
\renewcommand\onecolumngrid{
\do@columngrid{one}{\@ne}%
\def\set@footnotewidth{\onecolumngrid}
\def\footnoterule{\kern-6pt\hrule width 1.5in\kern6pt}%
}

\renewcommand\twocolumngrid{
        \def\footnoterule{
        \dimen@\skip\footins\divide\dimen@\thr@@
        \kern-\dimen@\hrule width.5in\kern\dimen@}
        \do@columngrid{mlt}{\tw@}
}%

\makeatother    

\usepackage{graphicx} 

\usepackage{amsmath,amsfonts,amssymb}
\usepackage{psfrag}
\usepackage{enumerate}
\usepackage{mathrsfs}
\usepackage{graphicx}
\usepackage{wrapfig}
\usepackage{xcolor}

\usepackage{amsthm}
\usepackage{xcolor}
\usepackage{esint}

\usepackage{comment}

\usepackage{array}
\usepackage{mathtools}
\usepackage{multirow}
\usepackage{varwidth}
\providecommand{\tabularnewline}{\\}
\newenvironment{cellvarwidth}[1][t]
    {\begin{varwidth}[#1]{\linewidth}}
    {\end{varwidth}}

\allowdisplaybreaks
 \newcommand{\be}{\begin{equation}}
\newcommand{\beq}{\begin{equation}}
 \newcommand{\ee}{\end{equation}}
 \newcommand{\bea}{\begin{align}}
 
 \newcommand{\eea}{\end{align}}

 \def\pg#1{\textcolor{cyan}{#1}}

\def\nn{\nonumber}

\global\long\def\mF{\mathcal{F}}%

\global\long\def\mJ{\mathcal{J}}%

\global\long\def\mO{\mathcal{O}}%

\global\long\def\e{\epsilon}%
 
\global\long\def\ra{\rightarrow}%

\global\long\def\avg#1{\left\langle #1\right\rangle }%

\global\long\def\f#1#2{\frac{#1}{#2}}%

\global\long\def\t{\theta}%

\global\long\def\b{\beta}%
 
\global\long\def\g{\gamma}%
 
\global\long\def\G{\Gamma}%

\global\long\def\r{\rho}%
 
\global\long\def\d{\delta}%

\global\long\def\N{\mathbb{N}}%

\global\long\def\Z{\mathbb{Z}}%
\global\long\def\E{\mathbb{E}}%
 
\global\long\def\R{\mathbb{R}}%

\global\long\def\w{\omega}%
 
\global\long\def\D{\Delta}%

\global\long\def\S{\Sigma}%

\global\long\def\app{\approx}%
\global\long\def\sgn{\text{sgn}}%
\global\long\def\lam{\lambda}%
\global\long\def\i{\text{i}}%

\newtheorem{conjecture}{Conjecture}

\newcommand{\ben}{\begin{enumerate}}
\newcommand{\een}{\end{enumerate}}
\newcommand{\bega}{\begin{gather}}
\newcommand{\eega}{\end{gather}}

\newcommand\field[1]{{\ensuremath{\mathbb{{#1}}}}}
\newcommand{\RR}{\field{R}}

\begin{document}

\preprint{MIT-CTP/5992}

\title{Probing Stringy Horizons with Pole-Skipping in Non-Maximal Chaotic Systems}
\author{Ping Gao} 
\email{gaoping@ucas.ac.cn}
\affiliation{Kavli Institute for Theoretical Sciences,
University of Chinese Academy of Sciences, Beijing 100190, China}

\author{Hong Liu}
\email{hong\_liu@mit.edu}
\affiliation{MIT Center for Theoretical Physics---a Leinweber Institute, 
Massachusetts Institute of Technology,
77 Massachusetts Ave.,  Cambridge, MA 02139 USA}

\begin{abstract}

In this paper, we study pole-skipping in non-maximally quantum chaotic systems. Using Rindler conformal field theories and the large-
$q$ SYK chain as illustrative examples, we argue that the pole-skipping points of few-body operators organize into trajectories in the complex frequency-momentum plane, with the leading trajectory encoding the quantum Lyapunov exponent. We further propose that these trajectories admit a natural interpretation as Regge trajectories of stringy excitations in a dual stringy black hole geometry. From this perspective, pole-skipping for an individual operator can be viewed as tracking the stringy horizon through the response of a single excitation. Our results suggest that pole-skipping reflects intrinsic properties of quantum chaotic systems and may be deeply connected to the structure of horizons in the stringy regime.

\end{abstract}

\maketitle

\section{Introduction}

Despite significant progress in recent years, characterizing quantum chaos in many-body systems remains a central challenge. Random matrix behavior~\cite{Boh84} provides one guiding framework, while the eigenstate thermalization hypothesis (ETH)~\cite{Sre94} offers another. In large-$N$ systems with many internal degrees of freedom, quantities such as the quantum Lyapunov exponent and the butterfly velocity---extracted from out-of-time-ordered correlators (OTOCs)---have emerged as key diagnostics of chaotic dynamics at short time scales~\cite{Shenker:2013pqa,Shenker:2013yza,Roberts:2014isa,Shenker:2014cwa,kitaevtalk15,Roberts:2014ifa}.  

For large-$N$ systems that saturate the maximal Lyapunov exponent, another manifestation of quantum chaos---known as pole-skipping~\cite{Grozdanov:2017ajz,Blake:2017ris}---has been identified. This phenomenon refers to the observation that the {\it thermal} retarded  function of the energy density in momentum space exhibits lines of poles and zeros in the complex frequency plane, which intersect at a special point $(\omega,k)_{\text{p.s.}}$ given by
\begin{equation}
(\omega,k)_{\text{p.s.}} = \i 2\pi/\beta (1,1/v_B),
\label{eq:388}
\end{equation}
where $2\pi/\beta$ (with $\b$ the inverse temperature) gives the maximal Lyapunov exponent $\lambda_{\text{max}}$, and $v_B$ is the butterfly velocity~\cite{Shenker:2013pqa,Shenker:2013yza}.
At this point, a would-be pole on the pole line is skipped due to an exact cancellation with a zero.

In maximally chaotic systems, the pole-skipping phenomenon can be understood as a consequence of the dual role played by the hydrodynamic mode associated with energy density: it simultaneously encodes energy conservation and the propagation of chaos~\cite{Blake:2017ris}. On the gravity side, this behavior has been traced to properties of metric perturbations near the horizon~\cite{Blake:2018leo} and further related to a class of near-horizon symmetries~\cite{KnyLiu24}. Pole-skipping in the energy-density channel has since been observed in all known maximally chaotic systems (see~\cite{BlaLiu21} for a general discussion), including holographic models dual to higher-derivative gravity theories~\cite{Grozdanov:2018kkt,Natsuume:2019xcy,Wu:2019esr,Baishya:2023mgz,Wang:2022mcq}. 

Interestingly, pole-skipping has also been observed across a wide class of operators and also in the {\it lower-half} frequency plane~\cite{Blake:2018leo,Blake:2019otz,Natsuume:2019xcy,Ahn:2019rnq,Wu:2019esr,Blake:2021hjj,Baishya:2023mgz,Grozdanov:2023txs,Wang:2022mcq,Ning:2023ggs,Ahn:2024gjh,Chua:2025vig}.
For example, for a scalar field, pole-skipping occurs only in the lower-half frequency plane, at $\omega = - \i 2\pi  n / \beta$ with $n = 1, 2, 3, \ldots$, with the corresponding momenta $k$ being model-dependent; for each $n$, there are $2n$ pole-skipping points in total~\cite{Blake:2019otz}.

Pole-skipping presents a range of open puzzles.
A central question concerns its status in non-maximally chaotic systems. It was shown in~\cite{Choi:2020tdj} that, in the large-$q$ SYK chain~\cite{Gu:2016oyy} model---which is non-maximal---the pole-skipping location of the energy density remains given by~\eqref{eq:388}, which appears to be unrelated to the non-maximal Lyapunov exponent~$\lambda$.  This raises the question of whether the connection between pole-skipping and chaos is restricted to maximal chaos.

In holographic systems, this issue is also closely tied to the nature of the black hole horizon in the stringy regime. In the Einstein regime, both the maximal Lyapunov exponent and the pole-skipping phenomenon can be understood in terms of the local boost symmetry near the horizon, whereas deviations from maximal chaos in the stringy regime have often been interpreted as indications that the horizon becomes ``fuzzy.''

In this paper, we clarify the nature of pole-skipping in general non-maximally chaotic systems and use it as a probe of the stringy horizon in holographic settings. More specifically, we argue that:
(i) pole-skipping should be viewed as a universal feature of chaotic systems;
(ii) it provides a mean to track both the leading bulk stringy Regge trajectory in a black-hole geometry and the associated Lyapunov exponent;
(iii) Rindler and black-hole horizons remain sharp even in the stringy regime, and their horizon symmetries persist.

\section{Pole-skipping of higher spin operators in a Rindler CFT} \label{sec:II}

To understand pole-skipping for AdS black holes in the stringy regime from the perspective of the dual CFT, one must analyze the corresponding phenomenon in the boundary theory on $\mathbb{R} \times S^{d-1}$ (or $\mathbb{R}^{1,d-1}$) in the large-$N$ limit, at finite coupling, and above the Hawking-Page temperature. This, however, is a highly challenging problem over which we currently lack full theoretical control. As a first step, we therefore consider a simpler, solvable setting that---as we will argue below---may already capture key features relevant to more general systems.

Specifically, we study a CFT on $\mathbb{R}^{1,d-1}$ in its vacuum state and focus on the right Rindler wedge---hereafter referred to as the Rindler CFT---which corresponds to the CFT on $\mathbb{R} \times H_{d-1}$ at inverse temperature~$\beta$, with $H_{d-1}$ denoting a $(d-1)$-dimensional hyperbolic space of radius~$\beta/2\pi$.
We set units such that~$\beta = 2\pi$. Four-point functions of the original CFT on $\mathbb{R}^{1,d-1}$ in the Regge regime are  mapped to  OTOCs of the Rindler CFT in a non-maximally chaotic regime~\cite{Roberts:2014ifa,Perlmutter:2016pkf,Mezei:2019dfv}.  
Since two-point functions in the Rindler CFT are fixed by conformal symmetry, studying pole-skipping in this setting provides an analytically tractable framework for probing this phenomenon in non-maximally chaotic systems and, as we will see, for investigating the bulk AdS-Rindler horizon in the stringy regime.

Pole-skipping of spin-1 and spin-2 (stress tensor) operators in the Rindler CFT has been previously studied in~\cite{Ahn:2020bks,Haehl:2019eae}. Here, we investigate pole-skipping for general higher-spin operators with spin~$J$, which, as we will see, uncovers a wealth of new and systematic information that cannot be captured from small-spin analyses or isolated higher-spin cases.\footnote{Pole-skipping of bulk higher-spin fields in AdS black holes has been previously studied in~\cite{Wang:2022mcq}. Owing to the highly involved nature of the bulk equations, only partial results were obtained. We will compare our findings with theirs below.}

Consider a primary higher-spin operator $\mathcal{O}_{\mu_1 \cdots \mu_J}$ with scaling dimension~$\Delta$ and spin~$J$. We focus on the component with all indices along the time direction, as this component provides the most complete information about pole-skipping.
Introducing a polarization vector $z^\mu$, the vacuum two-point function of the polarized operator $\mathcal{O}(x, z) = 
\mathcal{O}_{\mu_1 \cdots \mu_J} z^{\mu_1} \cdots z^{\mu_J}$ is given by
\begin{equation}
\langle \mathcal{O}(x, z)\mathcal{O}(x', z') \rangle
= \frac{[-z \cdot I(x - x') \cdot z']^{J}}{[(x - x')^{2}]^{\Delta}}
\label{eq:2.1}
\end{equation}
where $I^{\mu}{}_{\nu}(x) \equiv \delta^{\mu}{}_{\nu} - \frac{2x^{\mu}x_{\nu}}{x^{2}}$, and the overall normalization has been set to unity. When both $x$ and $x'$ lie in the Rindler wedge and are expressed in Rindler coordinates,~\eqref{eq:2.1} yields the thermal two-point function of the Rindler CFT. Fourier transforming this expression in Rindler coordinates allows us to obtain the corresponding retarded correlator in momentum space and identify the pole-skipping points. We relegate the details to Supplementary Material~\ref{app:Fourier} and present only the final results here.

With $\w$ and $k$ denoting the frequency for the Rindler time and the magnitude of the spatial momentum,\footnote{Although there is no translation symmetry along the radial direction in hyperbolic space, it is nevertheless possible to define $k$; see~\ref{app:Fourier}.} we find the pole-skipping points are located at\footnote{There are also a few pole-skipping points which location depend on the specific operator choice in a non-universal way, we will neglect them. See~\ref{app:Fourier}.}
\begin{equation}
(\w,k)_{p.s.}=\begin{cases}
\i(J-2n-1-s,\g-s) & n \leq J\\
\i(-n-1-s,\g+n-J-s) & n \geq J
\end{cases}
\label{eq:512-2}
\end{equation}
where $\g=\D-d/2$ and $n,s=0,1,2,\cdots$, together with another family obtained by taking $k \to -k$. We note the following features: 

\begin{enumerate}

\item The
highest pole-skipping points for each $J$ lies on the $(\Im \w,\Im k)$ plane at 
\begin{equation}
(\w,k)_{p.s.}^{\rm (highest)}=\i(J-1,\D-d/2)  \ .
\label{eq:2.47}
\end{equation}
For the stress tensor, with $J=2$ and $\D =d$, equation~\eqref{eq:2.47} can be related to \eqref{eq:388}, corresponding to the theory on $\RR^{1, d-1}$ at finite temperature, where $v_B =1/(d-1)$ by a shift $k\ra k+\i(d/2-1)$ \cite{Haehl:2019eae,Choi:2020tdj}. 

\item All these pole-skipping points lie at the Matsubara frequencies (recall $\b = 2 \pi$ here), 
\be
\w=2\pi \i (J-n)/\b \quad \text{with} \quad  n=1,2,3,\cdots  ,
\ee 
which agrees with the results of~\cite{Wang:2022mcq} for bulk higher-spin fields in AdS black holes.\footnote{Note that~\cite{Wang:2022mcq} does not obtain all the pole-skipping points for each $n$.} 

\item For each $J$, writing $\w = \i m$ with $m \leq J-1$, we find that it is natural to separate the pole-skipping points for each operator into two families based on the number $N_m$ of pole-skipping points for each $m$
\begin{equation}
N_{m}=\begin{cases}
2\left\lfloor \f{J-1-m}2\right\rfloor +2 & -(J-1)\leq m\leq J-1\\
2|m| & m \leq -J
\end{cases} \ .
\label{eq:513-2}
\end{equation}

The second group was motivated from the result of a scalar operator in holographic systems~\cite{Blake:2019otz} where the number of point-skipping points at each $\w_{m}$
are $2|m|$.  For $J=1$ and $J=2$, the first group contains one and three rows of pole-skipping points respectively, which are consistent with previous works \cite{Ahn:2020bks,Haehl:2019eae}.

\end{enumerate}

\begin{figure}
\begin{centering}
\includegraphics[width=8.5cm]{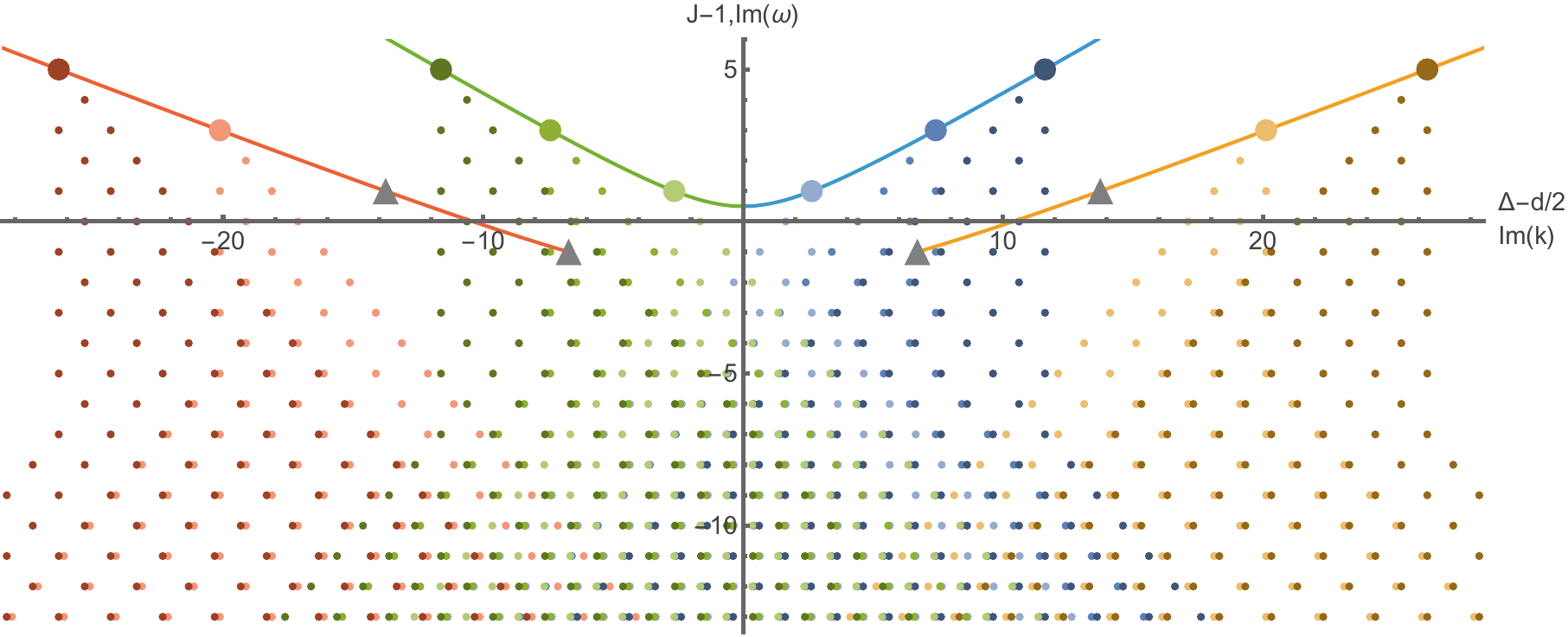}
\par\end{centering}
\caption{The overlay of Regge trajectories and the pole-skipping points. The
horizon axis is $\protect\D-d/2,\Im k$ and the vertical axis is $J-1,\Im\protect\w$.
The blue and green curve together is the leading Regge trajectory; The yellow and red curves are two symmetric subleading Regge
trajectories. Each dot
on the trajectories is a physical operator with even spin, which seeds a family
of pole-skipping points \eqref{eq:512-2} (smaller dots with variant colors) below it. Pole-skipping for odd spins likewise forms trajectories, but since they are not involved in the leading trajectory, we suppress them to avoid cluttering the plot. The triangles are missing or decoupling zeros on a subleading trajectory \cite{Homrich:2022cfq,Henriksson:2023cnh}, where no physical operator exists at an integer spin, and are irrelevant to pole-skipping. \label{fig:The-overlay-of}}
\end{figure}

Recall that in a large-$N$ CFT, operators can be organized into discretely spaced Regge trajectories on the $(\Delta, J)$ plane~\cite{Penedones:2010ue,Costa:2012cb,Caron-Huot:2017vep,Li:2017lmh,Brower:2006ea}.  Each trajectory consists of an infinite sequence of physical operators whose scaling dimensions $\Delta(J)$ (or equivalently, $\gamma(J) = \Delta(J) - \tfrac{d}{2}$) take discrete values at integer spin. In principle, the full analytic Regge trajectory $\gamma(J)$, or its inverse function $j(\mu)$~(with $\g=\i\mu$), can be reconstructed from these lattice points by analytic continuation in spin (see e.g.,~\cite{Homrich:2022cfq} for a discussion). Among all the trajectories, the leading Regge trajectory $j_{\rm leading} (\mu)$---defined as the uppermost curve in the $j(\mu)$ versus $\mu$ plane, and hence the one with the largest intercept $j(0)$---plays a particularly important role, as it governs the dominant behavior of conformal partial waves in the Regge limit of CFT correlators, much like its counterpart in flat-space S-matrix theory.
In the Rindler CFT, the Lyapunov exponent extracted from an 
OTOC---equivalently, from the four-point function of the original CFT in the Regge regime---is determined by the intercept of $j_{\rm leading} (\mu)$ at $\mu = 0$,
\begin{equation} \label{Lya}
\lambda = j_{\rm leading} (0) - 1 \ .
\end{equation}
The exponent $\lambda$ encapsulates the collective contribution from the exchange of an infinite tower of higher-spin operators.

From~\eqref{eq:2.47}, we see that the highest pole-skipping point for each operator lies precisely on its Regge trajectory (see FIG.~\ref{fig:The-overlay-of}). Therefore, knowing the pole-skipping points---which correspond to the locations of physical operators along their Regge trajectories---allows, in principle, the reconstruction of all the trajectories, including the leading  trajectory. This, in turn, enables us to extract all physical information encoded in the Regge data, including the non-maximal Lyapunov exponent~\eqref{Lya}.\footnote{More generally, one can define the velocity-dependent Lyapunov exponent (VDLE) $\lam(v)$ \cite{Khemani:2018sdn} for OTOC with large spacetime separation with $v=|\r/T|$. It is given by the saddle-pole transition on the leading Regge trajectory \cite{Mezei:2019dfv}. In the pole-dominant regime, VDLE is ballistic $\lam(v)=2\pi/\beta(1-v/v_B)$, where $v_B$ is the butterfly velocity. The pole-dominant regime can be determined by the pole-skipping of stress tensor at \eqref{eq:2.47} with $J=2,\D=d$. Therefore, all quantum chaotic data of OTOC can be extracted from pole-skipping.}
Thus, while the pole-skipping of the stress tensor or any single higher-spin operator may appear unrelated to the non-maximal Lyapunov exponent~$\lambda$, the collective set of pole-skipping points across all operators reconstructs the full Regge trajectories and hence determines~$\lambda$.


\section{Stringy horizons and pole-skipping for general theories} 

The above discussion is specific to the Rindler CFT, where we established a connection between pole-skipping and the non-maximal Lyapunov exponent through the bridge of CFT Regge trajectories. In more general cases---such as a large-$N$ CFT on $\mathbb{R}^{1,d-1}$ or $\mathbb{R} \times S^{d-1}$ at finite temperature and finite coupling (hence non-maximally chaotic)---such a bridge does not appear to be available. In these theories, pole-skipping need not be governed by the CFT Regge trajectories, and the Regge data need not determine the non-maximal Lyapunov exponent. 

We will now argue that the pole-skipping structure observed here reveals features of the horizon geometry in the stringy regime, which in turn suggests a natural generalization to broader classes of theories, where the connection between pole-skipping and the Lyapunov exponent is mediated by bulk stringy Regge trajectories.



Consider a large-$N$ CFT with a gravitational dual. The boundary Rindler horizon is dual to the AdS Rindler horizon, 
and the Rindler CFT is dual to the bulk theory in the corresponding AdS Rindler region.
 Vacuum four-point functions of the boundary CFT in the Regge regime map to high-energy AdS scattering in the corresponding bulk Regge limit~\cite{Brower:2006ea,Penedones:2010ue,Costa:2012cb}, effectively occurring across the AdS Rindler horizon.  
 
 In the semiclassical gravity limit---dual to the boundary CFT at infinite coupling---only the graviton is exchanged, corresponding to the exchange of the boundary stress tensor, and yielding the maximal Lyapunov exponent $\lambda_{\text{max}} = 1$. If one were to include the exchange of a hypothetical higher-spin particle of spin~$J$, the corresponding Lyapunov exponent would be $\lambda_J = J - 1$, although no such particles exist in the low-energy spectrum.
 
In the stringy regime, however, an infinite tower of bulk higher-spin string excitations appears, dual to higher-spin single-trace boundary operators whose dimensions become finite at finite coupling. {\it The Regge trajectories of these boundary operators coincide with those of the bulk stringy excitations.} In the corresponding AdS Rindler scattering process, the intermediate states include all such higher-spin exchanges, whose collective contribution yields a non-maximal Lyapunov exponent $\lambda < 1$.

Our discussion of pole-skipping in the Rindler CFT relies only on conformal symmetry and therefore extends to the boundary CFT at finite coupling, which is dual to the bulk theory in the stringy regime. In particular, the pole-skipping of a single-trace spin-$J$ boundary operator in the Rindler CFT maps to that of the corresponding bulk stringy field in the AdS Rindler region. Consequently, all the features discussed in the previous section translate directly into properties of bulk stringy fields relative to the AdS Rindler horizon:
\begin{enumerate}
\item From the conformal partial wave analysis, in the stringy regime, each higher-spin field still contributes individually to AdS Regge scattering with an exponent $\lambda_J = J - 1$. This ``field-dependent Lyapunov exponent'' is still captured by the highest pole-skipping point of the field.

\item 
The highest pole-skipping points of bulk stringy fields collectively trace out the Regge trajectories of {\it bulk} string excitations. 

\item The full analytic Regge trajectories of the bulk string theory can be reconstructed from the pole-skipping points via analytic continuation, including in particular the leading trajectory $j_{\rm leading}(\mu)$---which corresponds to the highest curve in the $(\Im k, \Im \w)$ plane---together with the associated non-maximal Lyapunov exponent $\lambda$.
\end{enumerate}
In semiclassical gravity, it has been argued that for $J = 2$ (metric perturbations), the highest pole-skipping point~\eqref{eq:2.47} can be attributed to certain horizon symmetries~\cite{KnyLiu24}. It appears reasonable to expect that this argument can be extended to fields of general spin.\footnote{We mention by passing that it appears natural to conjecture that the first line of~\eqref{eq:513-2} corresponds to ``descendants'' of the highest  pole-skipping points, arising from the same underlying horizon symmetries, while the second line originates from a different physical mechanism. 
} Therefore, equation~\eqref{eq:2.47} suggests that: 
\begin{enumerate}
\setcounter{enumi}{3}
\item The horizon symmetries underlying pole-skipping persist into the stringy regime for each individual bulk field.
\end{enumerate}

At strong coupling, where the bulk dynamics is described by classical gravity---including both Einstein and higher-derivative 
theories---there is strong evidence that the pole-skipping frequencies  $\omega = \tfrac{2 \pi i (J - n)}{\beta}, \; n=1,2,\cdots$ for each bulk field depend only on the near-horizon geometry~\cite{Blake:2018leo,Wang:2022mcq}, irrespective of whether the horizon corresponds to a black hole, an AdS Rindler patch, or other background details.\footnote{For the highest pole-skipping points, an explanation for this universality is their origin from horizon symmetries.} Thus we conjecture that {\it the statements in items~1-4 above apply  to a stringy black hole horizon as well.}

Now consider a boundary CFT on $\mathbb{R} \times S^{d-1}$ at a temperature $1/\b$ above the Hawking-Page transition and at finite coupling, dual to an eternal black hole in the stringy regime. An OTOC in the boundary CFT maps to a stringy Regge scattering process in the black hole geometry~\cite{Shenker:2014cwa}. 
Translating items~1-4 above into the boundary description then leads to the following {\it conjectures}:

\ben 

\item In a general non-maximally chaotic theory at a finite temperature, the highest pole-skipping points at imaginary Matsubara frequencies align along certain trajectories, among which there is 
a leading trajectory $\Im \w = \lambda (k)$, whose intercept $\lambda(0)$ gives the Lyapunov exponent. \label{conj1}

\item For systems with a bulk stringy dual, the trajectories of pole-skipping points correspond to the Regge trajectories of stringy excitations in the associated black hole geometry. 
\label{conj2}

\een

This picture---that in the stringy regime, at leading order in the $g_s$ expansion (the free-string limit), the horizon remains sharp for each individual higher-spin bulk stringy field---is also consistent with recent discussions in~\cite{GesLiu24,HerKud25}. Stringy fuzziness arises only in observables that involve summing over the full Hagedorn density of modes or that incorporate interactions.

In the Einstein regime, the Lyapunov exponent appears to be directly related to the horizon boost symmetry. However, when viewed within the broader stringy regime, this relation seems accidental: there, $\lambda = \tfrac{2\pi}{\beta}$ is determined by the stress tensor, which itself is fixed by the horizon symmetry. In the stringy regime, although the same horizon symmetries persist, the Lyapunov exponent emerges from the collective contribution of an infinite tower of higher-spin modes---each individually respecting the same symmetry---yet their sum is no longer directly governed by it. In other words, each discrete point on the stringy Regge trajectory reflects a horizon symmetry, but the intercept does not.

Given its close connection to horizon structure, pole-skipping should be viewed as a universal feature of quantum chaotic systems---one that is distinct from, and perhaps even more fundamental than, the Lyapunov exponent itself, since the Lyapunov exponent is merely one of several pieces of information extractable from it.

In the next section, we provide evidence for the above conjectures 
 using the large-$q$ SYK chain model, even though it is neither conformal nor possesses a notion of spin.


\section{Trajectories of pole-skipping for large-$q$ SYK chain}


We will now examine pole-skipping for  large-$q$ SYK chain~\cite{Gu:2016oyy}, which is believed to be dual to a ``string-type'' gravity theory. The SYK chain consists of $M$ sites labeled by $z=0,\cdots,M-1$ to form a spatial chain, and $N$ Majorana fermions $\psi^z_j$ ($j=1,\cdots,N$) on each site with an SYK Hamiltonian plus a nearest neighbor coupling term of similar type (see Supplementary Material~\ref{app:SYK} for details).  $q$ denotes the number of fermions in each interaction term. 
In the large-$q$ limit, this model admits a simple Liouville-like effective action~\cite{Choi:2020tdj}, allowing many quantities to be computed analytically. We set units such that the inverse temperature is $\beta = 2\pi$, and the Lyapunov exponent $\lambda_0 < 1$ can  be obtained explicitly~(see~\eqref{lyaS}).

Since it is a lattice model, the spatial momentum $p=2\pi k/M$ is discretized with $k\in\Z_M$. We will take the continuum limit $M\ra\infty$ and regard $p$ as a real number in the first Brillouin zone $p\in[-\pi,\pi]$. In the discussion of pole-skipping, we will analytically continue $p$ to complex values. At order $1/N$, pole-skipping in this model was analyzed in \cite{Choi:2020tdj} for the two-point function of the composite operator 
$\varepsilon_{z}(t) 
= \frac{1}{\i q} \sum_{i} \psi_{i}^{z}(t) \partial_{t}\psi_{i}^{z}(t)$.

\begin{figure}
\begin{centering}
\includegraphics[width=8.5cm]{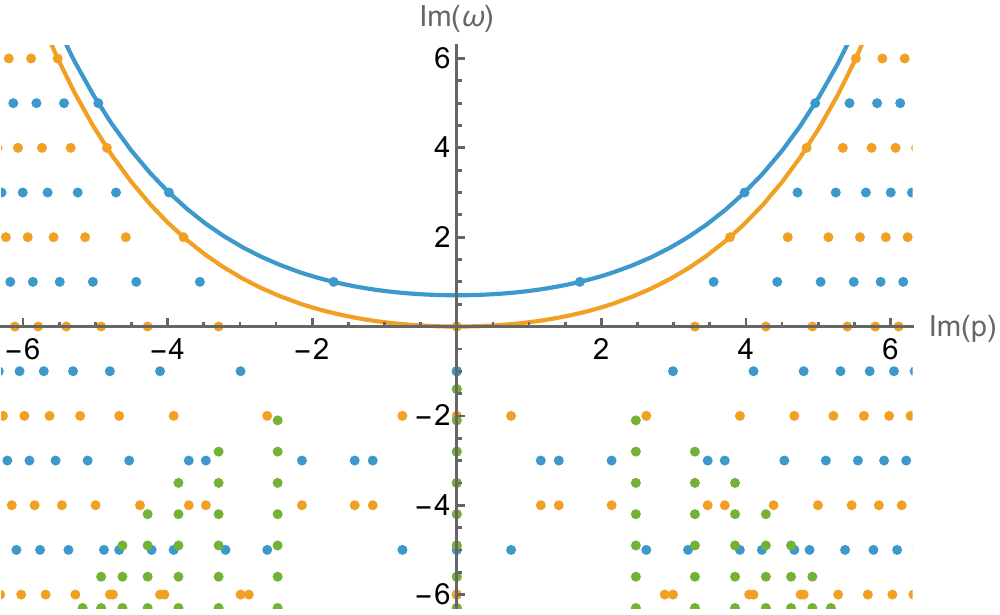}
\par\end{centering}
\caption{The pole-skipping points of the large $q$ SYK chain on the $(\Im \w,\Im p)$ plane with parameters $\eta=0.8$ and $\lam_0=0.7$. The blue and yellow dots are \eqref{eq:480-3} with odd and even $n$ respectively, and the green dots are \eqref{eq:459}. The Pole-skipping points without pure imaginary $p$ are not included in this plot. The blue and yellow curves are the two bounds in \eqref{eq:522} respectively. \label{fig:ps-sykchain}}
\end{figure}

Before discussing their results, we note that the two-point function of $\varepsilon_{z}$ corresponds to the time-ordered four-point function (TOC) of SYK fermions. Consequently, the associated pole-skipping structure should reduce to that of the intermediate operators exchanged in the four-point function, together with any additional ad hoc contributions arising from the specific definition of $\varepsilon_{z}$ itself. In other words, by examining the pole-skipping behavior of such composite operators, we can in principle extract the pole-skipping points of a large class of operators in the theory---namely, those bilinear in $\psi$.\footnote{In Supplementary Materials~\ref{app:II-ps}, we give a precise definition of composite operators in a generic theory and study their pole-skipping behavior in the Rindler CFT, where additional pole-skipping points are observed, potentially attributable to nonlocal intermediate operators. We leave elucidation of their origin to future work.} In~\cite{GaoLiu:sykch}, we show that the two families identified in~\cite{Choi:2020tdj} (shown in FIG.~\ref{fig:ps-sykchain} and described below) are the only ones that are universal and independent of the specific choice of composite operator. They can therefore be identified with the full universal set of pole-skipping points associated with bilinear operators. Note that, in contrast to the AdS-Rindler case, we can no longer separate the pole-skipping points associated with individual operators, as those operators are unknown. What we obtain instead is the full universal collection.

The first family of pole-skipping points lies at 
\begin{equation}
(\w,h(p))_{p.s.}=(\i n,\f n{\lam_{0}}+\f{1+(-)^{n}}2+2k+1) \label{eq:480-3}
\end{equation}
where $n\in\Z,~k= 0,1, \cdots$, $h(p)$ is a function of $p$ defined as
$h(h-1)/2=1+\f{\eta}2(\cos p-1)$,\footnote{We should choose the solution of $h\in(1,2]$ for $p\in[-\pi,\pi]$.}
and $\eta \in [0,1]$ is a constant which can be expressed in terms of the couplings of the theory. 
Note that all the points have imaginary Matsubara frequencies. 
For each $n$, there are infinitely many pole-skipping points as $k$ is unbounded from above. The special choice with $n=1$ and $k=0$ corresponds to the pole-skipping point (\ref{eq:388}) related to the maximal Lyapunov exponent.

Since SYK chain does not have conformal symmetry nor even Lorentzian symmetry, there does not exist a notion of ``higher spin" operator in this model. However, from FIG.~\ref{fig:ps-sykchain}, we see that, for this family of pole-skipping, all pole-skipping points have $\Im \w$ bounded from above by
\begin{equation}
\Im\w\leq \begin{cases}
\lam_{0}(h(p)-1), & \Im \w=n\text{ odd}\\
\lam_{0}(h(p)-2), & \Im \w=n\text{ even}
\end{cases}\ . \label{eq:522}
\end{equation}
Each upper bound is saturated by infinitely many pole-skipping points with nonnegative integer $n\geq 0$. We thus identify the first line of \eqref{eq:522} as giving the leading trajectory\footnote{$j_{\text{leading}}(p) -1 = (h(p)-1)\lam_0$ which is also precisely the momentum-dependent Lyapunov exponent  discussed by~\cite{Choi:2020tdj}.}
\be\label{ylea}
j_{\text{leading}}(p)=(h(p)-1)\lam_0+1 ,
\ee
which indeed recovers the Lyapunov exponent with $j_{\text{leading}}(0)-1 = \lam_0$ ($h(0) =2$). 
The large-$q$ SYK chain  thus verifies beautifully the conjecture \ref{conj1}.  Note that we have essentially extracted the Lyapunov exponent from the TOC. Although we cannot separate the pole-skipping points associated with individual operators, we have nevertheless extracted the leading trajectory, which can be interpreted as the leading Regge trajectory of the ``string theory'' dual to the SYK chain. It would be interesting to construct each individual operator on this trajectory and check if its Regge scattering contribution  matches with the conjecture \ref{conj2}.

There is a second family of pole-skipping points at
\begin{equation}
(\w,p)_{p.s.}=\left(-\i\lam_{0}n,\pm\arccos\left[1+\f{k(k-1)-2}{\eta}\right]\right)\label{eq:459}
\end{equation}
where $k=1,\cdots,n,$ and $n=1,2,\cdots$. In contrast with the first family, this family exhibits two new features: (i) Their locations are not at the imaginary Matsubara frequencies of the physical temperature, but instead at the Matsubara frequencies of the so-called fake temperature $T_{\mathrm{f.d.}}=\lambda_0/(2\pi)$~\cite{Lin:2023trc}.
(ii) They appear only in the lower-half frequency plane, and the total number of pole-skipping points at fixed $n$ is $2n$.
The second feature suggests that these are the pole-skipping points associated with a (bulk stringy) ``scalar'' operator, while the first feature indicates that, unlike other operators that ``experience'' the physical temperature, this operator effectively ``sees''  the fake temperature.

\section{Conclusion}

Using Rindler CFTs and the large-$q$ SYK chain as illustrative examples, we have argued that in generic non-maximally chaotic systems, the pole-skipping points of few-body operators organize into trajectories in the complex frequency-momentum plane. The leading trajectory encodes the quantum Lyapunov exponent. We further proposed that these trajectories admit a natural interpretation as Regge trajectories of stringy modes in a dual stringy black hole geometry. From this viewpoint, pole-skipping for an individual operator can be seen as ``tracking'' the stringy horizon via the response of a single stringy excitation.

We also emphasized the power of analyzing pole-skipping for generic composite operators. This approach allowed us to extract the universal set of pole-skipping points associated with essentially all operators, without requiring a detailed analysis of each individual field. Notably, this information alone was sufficient to reconstruct the leading trajectory that determines the Lyapunov exponent. Moreover, such an analysis could potentially reveal pole-skipping associated with nonlocal operators.

Overall, our results suggest that pole-skipping reflects intrinsic and universal features of quantum chaotic systems. Developing a more microscopic understanding of its origin in generic chaotic models would be highly desirable, both to clarify its connection to horizon structure and to place our conjectures on firmer ground. In particular, Ref.~\cite{GesLiu24} proposed defining stringy horizons in terms of half-sided modular inclusions in the boundary theory, which capture quantum ergodicity and chaoticity at the algebraic level. It is tempting to speculate that the emergence of pole-skipping is rooted in this underlying modular structure.

Progress along these lines would open new avenues for understanding quantum chaos and black hole dynamics, and yield fresh insights into the structure of horizons in the stringy regime.

\section*{Acknowledgements} We would like to thank Yue-Zhou Li and Douglas Stanford for stimulating and helpful discussions. PG is grateful for the NHETC of Rutgers University and the Institute of Advanced Studies, where part of this work was done. PG is supported by the Fundamental Research Funds for the Central Universities. HL is supported by the Office of High Energy Physics of U.S. Department of Energy under grant Contract Number  DE-SC0012567 and DE-SC0020360 (MIT contract \# 578218), and grant \#63670 from the John Templeton Foundation.

\bibliographystyle{JHEP.bst}
\bibliography{main.bib}

\clearpage

\onecolumngrid

\begin{center}
\textbf{\large Supplementary Materials}
\end{center}
\setcounter{equation}{0}
\setcounter{figure}{0}
\setcounter{table}{0}

\setcounter{section}{0}

\makeatletter
\renewcommand{\thesection}{S-\Roman{section}}
\renewcommand{\theequation}{S\arabic{equation}}
\renewcommand{\thefigure}{S\arabic{figure}}
\renewcommand{\thetable}{S\arabic{table}}

\section{Pole-skipping of a spin $J$ operator} \label{app:Fourier}

\subsection{Mode expansion on hyperbolic space for a single block\label{subsec:Mode-expansion-on}}

The Minkowski coordinates $x = (t, y, \vec{x}_\perp)$ and the Rindler coordinates $(T, \rho, \vec{x}_\perp)$ are related by
\begin{equation}
t=e^\r\sinh T,\quad y=e^\r \cosh T, 
\label{eq:9}
\end{equation}
and the Minkowski metric can be written in Rindler coordinates as
\begin{equation}
ds^{2} = -e^{2\r}dT^{2}+ \left( d\r^{2}+d\vec{x}_{\perp}^{2}\right)
\equiv e^{2 \rho} ds^2_{\RR \times H^{d-1}} 
\label{eq:106}
\end{equation}
where $ds^{2}_{\mathbb{R} \times H^{d-1}}$ denotes the metric of the Rindler CFT, with $\rho$ the radial coordinate on the hyperbolic space $H^{d-1}$.
Without loss of generality, from now on, we take 
\be 
x=(T,\r,\vec x_{\perp}),\quad x'=(0,\r',0) \label{eq5}
\ee
in terms of which~\eqref{eq:2.1} becomes
\begin{equation}
G^{\pm}_{\D,J}=\f{((\cosh(\r-\r')+\f 12x_{\perp}^{2}e^{-\r-\r'})\cosh T_\pm-1)^{J}}{(\cosh(\r-\r')+\f 12 x_{\perp}^{2}e^{-\r-\r'}-\cosh T_\pm)^{\D+J}}, \label{eq:484-2}
\end{equation}
where $T_\pm=T\pm\i\e$ with infinitesimal imaginary part to denote different Wightman functions.

It is convenient to expand $G^\pm_{\Delta,J}$ in Fourier modes $(\omega, p_\perp)$ along the $T$ and $\vec{x}_{\perp}$ directions, together with the normalized eigenfunctions $f_{k,p_\perp}(\rho)$ of the Laplacian on hyperbolic space \cite{Ohya:2016gto}
\begin{align}\label{fdef}
f_{k,p_{\perp}}(\r)=f_{-k,p_{\perp}}(\r) & =\sqrt{\f{4k\sinh\pi k}{\pi}}e^{\r(d-2)/2}K_{\i k}(|p_{\perp}|e^{\r})\\
\int_{-\infty}^{+\infty}d\r e^{-\r(d-2)}f_{k,p_{\perp}}(\r)f_{k',p_{\perp}}(\r) & =2\pi(\d(k-k')+\d(k+k'))
\end{align}
where $k$ denotes the quantum number associated with the radial coordinate~$\rho$. It follows that 
\begin{align}
G^\pm_{\D,J}=&\int_{-\infty}^{+\infty}\f{d\w dk}{(2\pi)^{2}}\int\f{d^{d-2}p_{\perp}}{(2\pi)^{d-2}}e^{-\i\w T+\i p_{\perp}\cdot x_{\perp}}\tilde{G}^\pm_{\D,J}(\w,k)f_{k,p_{\perp}}(\r)f_{-k,p_{\perp}}(\r')
\end{align}
The function $\tilde{G}^\pm_{\Delta,J}(\omega,k)$ gives the ``momentum-space'' correlator, which, interestingly, admits a two-dimensional Fourier representation over $T$ and an auxiliary spatial coordinate~$x$ that effectively encodes the $\rho$-direction:
\begin{align}
\tilde{G}^\pm_{\D,J}(\w,k) & =\f{(2\pi)^{\f{d-2}2}\G(\D+1-\f d2)}2\int_{-\infty}^{+\infty}dxdTe^{\i\w T-\i kx}H_{\D,J}(T_{\pm},x)\label{eq:493-1-1}\\
H_{\D,J}(T,x) & \equiv\f{(\cosh x\cosh T-1)^{J}{}_{2}F_{1}(-J,\f d2-1,\D,\f{\sinh^{2}T}{\cosh T\cosh x-1})}{\G(\D)(\cosh x-\cosh T)^{\D+J+1-d/2}}\label{eq:494-1-1}
\end{align}
Note that $H_{\D,J}(T,x)$ has an (auxiliary) ``lightcone'' structure at $T=\pm x$ and
the denominator has a branch cut along $T=\pm x$ to $T\ra\pm\infty$. The retarded function in ``momentum space'' is given by the Fourier transformation of the difference of two Wightman functions in \eqref{eq:493-1-1} with integration only in the future wedge
\be
\tilde{G}^{R}_{\D,J}(\w,k) \propto \int_{T>|x|}dxdTe^{\i\w T-\i kx}(H_{\D,J}(T_-,x)-H_{\D,J}(T_+,x))\label{eq9}
\ee

The derivation of \eqref{eq:493-1-1} is as follows. For the Wightman functions $G^\pm_{\D,J}$ in (\ref{eq:484-2}), let us first continue $T_\pm \ra T\pm \i (\pi-\d)$ to compute the Fourier transformation and continue it back in the end. In terms of an integral, the Wightman functions after continuation are
\begin{equation}
G^\pm_{\D,J}=e^{\pm\i\pi J\sgn T}\int_{-\i\infty}^{\i\infty}\f{dt}{2\pi\i}\f{(t\cosh T+1)^{J}}{(t+\cosh T)^{\D+J}}\f 1{\cosh(\r-\r')+\f 12x_{\perp}^{2}e^{-\r-\r'}-t}
\end{equation}
where the contour integral of $t$ can be deformed to the right to take the residue. This form separates
the $T$ coordinate from all spatial coordinates. Let us take the
ansatz
\begin{equation}
\f 1{\cosh(\r-\r')+\f 12x_{\perp}^{2}e^{-\r-\r'}-t}=\int_{-\infty}^{+\infty}\f{dk}{2\pi}\int\f{d^{d-2}p_{\perp}}{(2\pi)^{d-2}}e^{\i p_{\perp}\cdot x_{\perp}}\tilde{g}(t,k)f_{k,p_{\perp}}(\r)f_{-k,p_{\perp}}(\r')
\end{equation}
where $\tilde{g}(t,k)=\tilde{g}(t,-k)$.
It is straightforward
to compute\footnote{This calculus is similar to \cite{Ohya:2016gto}.}
\begin{align}
&2\tilde{g}(t,k)f_{k,p_{\perp}}(\r') =\int_{-\infty}^{+\infty}d\r e^{-\r(d-2)}\int d^{d-2}x_{\perp}\f{f_{k,p_{\perp}}(\r')e^{-\i p_{\perp}\cdot x_{\perp}}}{\cosh(\r-\r')+\f 12x_{\perp}^{2}e^{-\r-\r'}-t}\nonumber \\
=&\int_{0}^{\infty}dze^{zt}\int_{-\infty}^{+\infty}d\r e^{-\r(d-2)}f_{k,p_{\perp}}(\r)e^{-z\cosh(\r-\r')}\int d^{d-2}x_{\perp}e^{-\i p_{\perp}\cdot x_{\perp}-\f z2x_{\perp}^{2}e^{-\r-\r'}}\nonumber \\
=&\sqrt{\f{4k\sinh\pi k}{\pi}}\int_{0}^{\infty}dze^{zt}\int_{0}^{\infty}\f{dy}{y^{d/2}}K_{\i k}(|p_{\perp}|y)e^{-z(y^{2}+y'{}^{2})/(2yy')}\left(\f{2\pi yy'}z\right)^{\f{d-2}2}e^{-\f{p_{\perp}^{2}}{2z}yy'}\nonumber \\
=&2f_{k,p_{\perp}}(\r')\int_{0}^{\infty}dze^{zt}\left(\f{2\pi}z\right)^{\f{d-2}2}K_{\i k}(z)\nonumber \\
=&f_{k,p_{\perp}}(\r')\int_{0}^{\infty}dze^{zt}\left(\f{2\pi}z\right)^{\f{d-2}2}\int_{-\infty}^{+\infty}dxe^{-z\cosh x-\i kx}\nonumber \\
=&f_{k,p_{\perp}}(\r')\int_{-\infty}^{+\infty}dx\f{\G(2-d/2)(2\pi)^{\f{d-2}2}}{(\cosh x-t)^{2-d/2}}e^{-\i kx}\label{eq:488-3-1}
\end{align}
where in the third line we changed variables $y=e^{\r}$ and $y'=e^{\r'}$,
in the fourth line we used the identity
\begin{equation}
\int_{0}^{\infty}\f{dy}yK_{\i k}(|p_{\perp}|y)e^{-z(y^{2}+y'{}^{2})/(2yy')-\f{p_{\perp}^{2}}{2z}yy'}=2K_{\i k}(z)K_{\i k}(|p_{\perp}|y')
\end{equation}
and in the fifth line we used an integral representation of $K_{\i k}(z)$.
From (\ref{eq:488-3-1}), we can read out
\begin{equation}
\tilde{g}(t,k)=\f 12\int_{-\infty}^{+\infty}dx\f{\G(2-d/2)(2\pi)^{\f{d-2}2}}{(\cosh x-t)^{2-d/2}}e^{-\i kx}
\end{equation}
which leads to
\begin{equation}
e^{\pm\w\pi}\tilde{G}^\pm_{\D,J}(\w,k)=\f 12\G(2-d/2)(2\pi)^{\f{d-2}2}\int_{-\infty}^{+\infty}dxdT\int_{-\i\infty}^{\i\infty}\f{dt}{2\pi\i}\f{(t\cosh T+1)^{J}}{(t+\cosh T)^{\D+J}}\f{e^{\i\w T-\i kx}e^{\mp\i\pi J\sgn T}}{(\cosh x-t)^{2-d/2}}
\end{equation}
Assuming $2-d/2$ is a generic number, the $t$ integral can be deformed
to the right and circles around the branch cut along $[\cosh x,+\infty)$.
Shifting $t\ra\cosh x+t$ and taking the phase $e^{\pm\i\pi(2-d/2)}$
appropriately, the $t$ integral becomes
\begin{align}
 & \sin\pi(2-d/2)\int_{0}^{\infty}\f{dt}{\pi}\f{(t\cosh T+\cosh x\cosh T+1)^{J}}{(t+\cosh T+\cosh x)^{\D+J}t^{2-d/2}}\nonumber \\
= & \f{\G(\D+1-\f d2)}{\G(2-d/2)\G(\D)}\f{(\cosh x\cosh T+1)^{J}}{(\cosh x+\cosh T)^{\D+J+1-d/2}}{}_{2}F_{1}(-J,\f d2-1,\D,-\f{\sinh^{2}T}{1+\cosh T\cosh x})
\end{align}
which leads to \eqref{eq:493-1-1} after the continuation $T\ra T\mp\i(\pi-\d)$ back to $T_{\pm}$. 

\subsection{Pole-skipping analysis} \label{app:ps-J}

For non-negative integer $J$, the hypergeometric function in $H_{\D,J}$
is a polynomial of order $J$. We can expand it and complete the Fourier
transformation term by term. The retarded function is restricted to
$T>|x|$ and take difference between two $H_{\D,J}$ with $T=T_{\pm}$,
which introduces a prefector of $e^{\pm\i\pi(\D+J+1-d/2)}$ respectively
due to the denominator of (\ref{eq:494-1-1}). Therefore, the momentum
space retarded function
\begin{equation}
\tilde G_{\D,J}^{R}(\w,k)=c_{\D,J}\int_{T>|x|}dxdTe^{\i\w T-\i kx}\f{(\cosh x\cosh T-1)^{J}{}_{2}F_{1}(-J,\f d2-1,\D,\f{\sinh^{2}T}{\cosh T\cosh x-1})}{(\cosh T-\cosh x)^{\D+J+1-d/2}}\label{eq:ps12}
\end{equation}
where $c_{\D,J}$ is a constant. Expanding the hypergeometric function
in series and $\sinh T$ in powers of $e^{-T}$, the integrand of
(\ref{eq:ps12}) becomes
\begin{equation}
\sum_{n=0}^{J}\sum_{s=0}^{2n}\f{(2n)!(-J)_{n}(d/2-1)_{n}(-)^{s}}{s!(2n-s)!4^{n}n!(\D)_{n}}e^{-2(n-s)T}\f{(\cosh x\cosh T-1)^{J-n}}{(\cosh T-\cosh x)^{\D+J+1-d/2}}\label{eq:ps13}
\end{equation}
The power of $e^{-T}$ simply shifts the frequency $\w\ra\w-2\i(n-s)$
for the Fourier transformation of the last term in (\ref{eq:ps13}).
Introducing the lightcone coordinates and the conjugate momenta
\begin{equation}
x_{\pm}\equiv(T\pm x)/2,\quad p_{\pm}\equiv\w\pm k \label{s14}
\end{equation}
we have
\begin{align}
 & \tilde{g}_{n}(\w,k)\equiv\int_{T>|x|}dxdTe^{\i\w T-\i kx}\f{(\cosh x\cosh T-1)^{J-n}}{(\cosh T-\cosh x)^{\D+J+1-d/2}}\nonumber \\
= & 2\int_{0}^{+\infty}dx_{\pm}e^{\i p_{-}x_{+}+\i p_{+}x_{-}}\f{(\sinh^{2}x_{+}+\sinh^{2}x_{-})^{J-n}}{(2\sinh x_{+}\sinh x_{-})^{\D+J+1-d/2}}\nonumber \\
= & \f 1{2^{\D+J-d/2}}\sum_{m=0}^{J-n}\f{(J-n)!}{m!(J-n-m)!}\int_{0}^{+\infty}\f{dx_{\pm}e^{\i p_{-}x_{+}+\i p_{+}x_{-}}}{(\sinh x_{+})^{\D+J+1-d/2-2m}(\sinh x_{-})^{\D-J+1-d/2+2n+2m}}\nonumber \\
= & \sum_{m=0}^{J-n}\frac{(J-n)!\Gamma\left(\frac{d}{2}-J+2m-\Delta\right)\Gamma\left(\frac{d}{2}+J-2(m+n)-\Delta\right)\Gamma\left(\frac{-d/2+J-2m+\Delta+1-\i p_{-}}{2}\right)\Gamma\left(\frac{-d/2-J+2(m+n)+\Delta-\i p_{+}+1}{2}\right)}{2^{J+d/2-\D-2n}m!(J-n-m)!\Gamma\left(\frac{d/2-J+2m-\Delta+1-\i p_{-}}{2}\right)\Gamma\left(\frac{d/2+J-2(m+n)-\Delta-\i p_{+}+1}{2}\right)}
\end{align}
It follows that 
\begin{align}
\tilde G_{\D,J}^{R}(\w,k)= & c_{\D,J}\sum_{n=0}^{J}\sum_{s=0}^{2n}\sum_{m=0}^{J-n}\f{(2n)!(-J)_{n}(d/2-1)_{n}(J-n)!(-)^{s}\Gamma\left(\frac{d}{2}-J+2m-\Delta\right)\Gamma\left(\frac{d}{2}+J-2(m+n)-\Delta\right)}{2^{J+d/2-\D}s!(2n-s)!n!(\D)_{n}m!(J-n-m)!}\nonumber \\
 & \times\frac{\Gamma\left(\frac{-d/2+J-2(m+n-s)+\Delta+1-\i p_{-}}{2}\right)\Gamma\left(\frac{-d/2-J+2(m+s)+\Delta-\i p_{+}+1}{2}\right)}{\Gamma\left(\frac{d/2-J+2(m-n+s)-\Delta+1-\i p_{-}}{2}\right)\Gamma\left(\frac{d/2+J-2(m+2n-s)-\Delta-\i p_{+}+1}{2}\right)}\label{eq:ps16}
\end{align}
where we have shifted $\w\ra\w-2\i(n-s)$ in (\ref{eq:ps16}). 

The pole lines are given by the gamma functions in the numerator of pole-skipping of the second line of (\ref{eq:ps16}), and the zero lines are due to either the gamma functions in the denominator of each term or their linear combination from the first line of (\ref{eq:ps16}). We will focus on the pole-skipping from the zeros of the first type because those of the second type do not have a universal pattern as we change $\D$ and $J$. Since \eqref{eq:ps16} is symmetric in $p_{-}$
and $p_{+}$, we can just check the pole lines of $p_{-}$ and zero
lines of $p_{+}$. The pole lines are due to the gamma function in
the numerator at nonpositive integers
\begin{equation}
\i p_{-}=-\frac{d}{2}+\Delta+J+2l_{1}-2m-2n+2s+1,\quad l_{1}\in\N\label{eq:ps17}
\end{equation}
The zero lines are due to the gamma function in the denominator at
nonpositive integers
\begin{equation}
\i p_{+}=\frac{d}{2}-\Delta+J+2l_{2}-2m-4n+2s+1,\quad l_{2}\in\N
\end{equation}
There are many terms in (\ref{eq:ps16}) giving the same pole line
and zero line with fixed $i_{1}=l_{1}-m-n+s$ and $i_{2}=l_{2}-m-2n+s$.
The range of $i_{1,2}$ are bounded by
\begin{equation}
i_{1}\geq s-(m+n),\quad i_{2}\geq s-(m+n)-n\label{eq:ps20}
\end{equation}
The pole-skipping points are at the crossing of the pole line and
zero line such that all $0\leq n\leq J$, $0\leq s\leq2n$, and $0\leq m\leq J-n$
satisfying the first inequality of (\ref{eq:ps20}) for a fixed $i_{1}$
must also satisfy the second inequality of (\ref{eq:ps20}), which
determines the lower bound of $i_{2}$ in terms of $i_{1}$. In other
words, all terms giving the same pole line must also have a zero line
to cancel it at the pole-skipping point, which are located on the
$(\Im\w,\Im k)$ plane at
\begin{equation}
(\Im\w,\Im k)_{p.s.}=(-J-i_{1}-i_{2}-1,\Delta-\frac{d}{2}+i_{1}-i_{2})\label{eq:ps19}
\end{equation}
It is straightforward to work out the lower bound of $i_{2}$ in terms
of $i_{1}$
\begin{equation}
i_{2}\geq\begin{cases}
i_{1} & 0\geq i_{1}\geq-J\\
0 & i_{1}\geq0
\end{cases}
\end{equation}
Taking $i_1=n-J$ and $i_2=i_1+s$ in the first case, and $i_1=n-J$ and $i_2=s$ in the second case, it follows that the pole-skipping points are two-piece distributed
\begin{equation}
(\Im\w,\Im k)_{p.s.}=\begin{cases}
(J-2n-1-s,\Delta-\frac{d}{2}-s) & 0\leq n\leq J\\
(-n-s-1,\Delta-\frac{d}{2}+n-J-s) & n\geq J
\end{cases},\quad s\in\N\label{eq:ps22}
\end{equation}
Since the pole-skipping points are $p_{\pm}$ symmetric, we should
also include the reflected ones with $k\ra-k$ in \ref{eq:ps22}.
A notable feature of these pole-skipping points is that their frequencies
are all at imaginary Matsubara frequencies and the highest one is
at $\Im\w=J-1$, which is consistent with bulk computations.

\begin{figure}
\begin{centering}
\includegraphics[width=7cm]{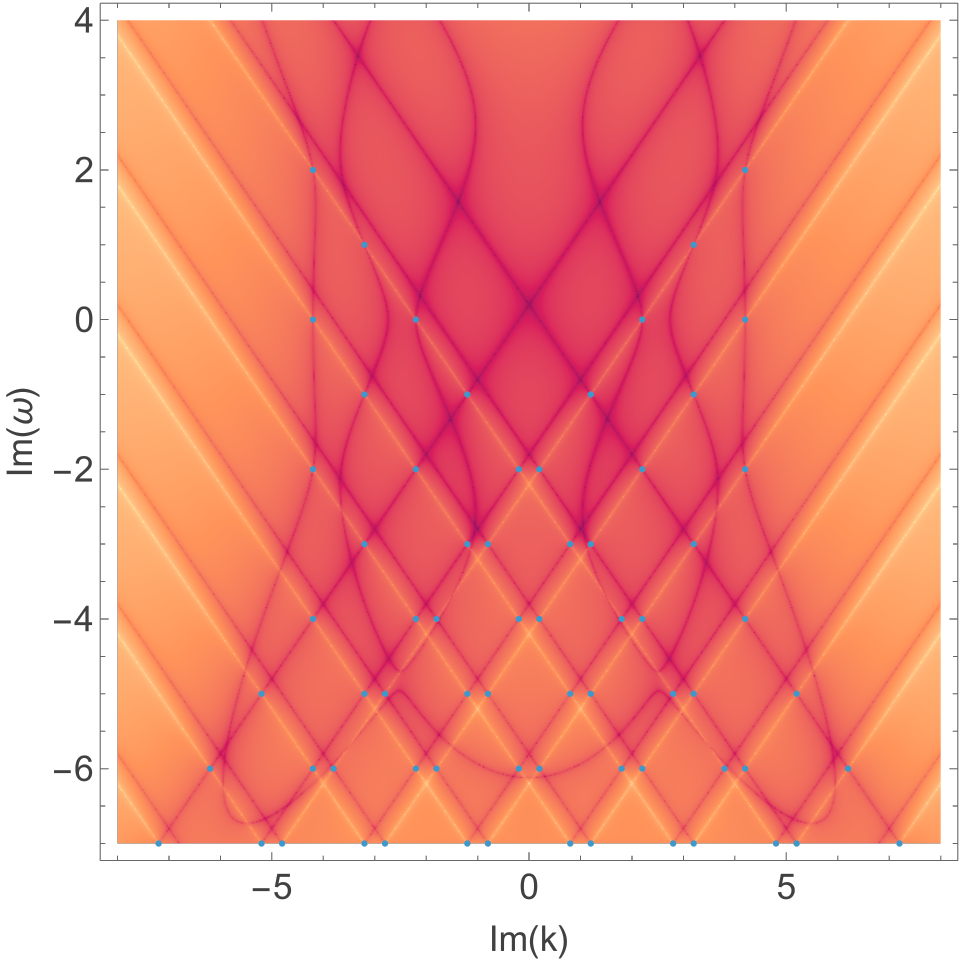}
\par\end{centering}
\caption{Pole-skipping of spin $J=3$ with parameter
$d=5$ and $\D=6.7$. The bright lines are pole lines, and
the dark lines are zero lines. The blue dots are the pole-skipping
points (\ref{eq:512-2}), which are at the intersection of a pair of pole line and zero line. There
are also a few sporadic crossings giving non-universal pole-skipping points other than the blue points. \label{fig:Pole-skipping-of-spin}}

\end{figure}

We draw these pole-skipping points for $J=3$ in FIG. \ref{fig:Pole-skipping-of-spin}.
It is clear that these pole-skipping points are at the intersection
of a pair of pole line and zero line. There are a few more sporadic pole-skipping
points in the figure, which corresponds to additional zeros from the linear combination of terms in (\ref{eq:ps16}) that shares the same pole
line. These pole-skipping points have $\Im\w$ not at the imaginary
Matsubara frequencies, and their locations vary in a non-universal way as we change $\D$ and $J$.


\section{Pole-skipping of composite operators in a Rindler CFT} \label{app:II-ps}

\subsection{Composite operators in a general theory}

For a general theory, consider bilinear composite operators of the form 
\begin{equation}
\mO_{W}(x)\equiv\int_{B_{\e}}dyf_{W}(y)W(x+y/2)W(x-y/2), 
\label{eq:436-1}
\end{equation}
where $W$ is some few-body operator, $B_{\e}$ is a $O(\e)\ll 1$ size small spacetime region around origin, and $f_{W}(y)$
is a smear function such that the integral is finite.\footnote{It is necessary to add infinitesimal imaginary time to avoid the universal
UV singualrities and make this integral well-defined. If $W$ is hermitian, we can construct a hermitian $\mO_{W}(x)$ by choosing $f_{W}(y)=f_{W}(-y)^{*}$.} A two-point function of such composite operators is essentially a time-ordered four-point function (TOC) 
\begin{align} 
&\avg{\mO_W(x)\mO_V(x')}=\int_{B_{\e}}dy dy' f_{W}(y)f_{V}(y')\avg{W(x+\f y 2)W(x-\f y 2)V(x'+\f {y'} 2)V(x'-\f {y'} 2)} , 
\label{eq:2.3}
\end{align}
where the ordering of $\mO_W$ and $\mO_V$ is given by appropriate $\i \d$-prescription for their imaginary time components. The retarded function for the composite operator is defined as
\be 
G^R_{\mO_W\mO_V}(x)=\i \t(x^0) \avg{[\mO_W(x), \mO_V(0)]}
\ee 
where we have used translation symmetry in $x^0$. Pole-skipping points of $G^R_{\mO_W\mO_V}$ in momentum space 
should reduce to those of the intermediate operators exchanged for the 4-point function plus possible ad hoc ones coming from the definition of $\mO_W, \mO_V$ themselves. We could thus extract the pole-skipping of the operators in the theory without knowing explicitly the operator spectrum or their two-point functions by extracting the universal part of the pole-skipping points from $G^R_{\mO_W\mO_V}$. In the following, we apply this definition to Rindler CFT and analyze the pole-skipping of a generic composite operator. Besides the pole-skipping points of each higher spin $J$ operator \eqref{eq:512-2}, there are exceptional pole-skipping points related to Regge trajectories that can be extracted in this way.

\subsection{Barnes integral representation of the momentum spacetime retarded function}

For a four-point function in a $d>2$ dimensional CFT, we can decompose it in terms of partial wave expansion of principal series representations $(\D,J)$ with conformal weight
$\D=d/2+\i\mu$ with $\mu\in\R$ and spin $J\in\N$ \cite{Dolan:2011dv,Kravchuk:2018htv,Karateev:2018oml,Gadde:2017sjg,Costa:2012cb}
\begin{align}
&\avg{\phi_{1}(x_{1})\cdots\phi_{4}(x_{4})} =\sum_{J=0}^{\infty}\int_{\f d2-\i\infty}^{\f d2+\i\infty}\f{d\D}{2\pi\i}a(\D,J)F_{\D,J}^{\D_{i}}(x_{i})\label{eq:1}\\
&F_{\D,J}^{\D_{i}}(x_{i}) =\f {\left(x_{24}^{2}/x_{14}^{2}\right)^{(\D_{1}-\D_{2})/2}\left(x_{14}^{2}/x_{13}^{2}\right)^{(\D_{3}-\D_{4})/2}}{(x_{12}^{2})^{(\D_{1}+\D_{2})/2}(x_{34}^{2})^{(\D_{3}+\D_{4})/2}}\mF_{\D,J}^{\D_{i}}(u,v) \label{eq:4.2}
\end{align}
where the cross ratios $u$ and $v$ are defined as 
\begin{equation}
u=\f{x_{12}^{2}x_{34}^{2}}{x_{13}^{2}x_{24}^{2}},\quad v=\f{x_{14}^{2}x_{23}^{2}} {x_{13}^{2}x_{24}^{2}} \label{2.7}
\end{equation}
Here $\mF_{\D,J}^{\D_{i}}$ is a conformal block completely determined by conformal symmetry \cite{Dolan:2011dv}
\begin{align}
\mF_{\D,J}^{\D_{i}}(u,v) & =\g_{\D,J}^{\D_{i}}G_{\D,J}^{\D_{i}}(u,v)+\g_{d-\D,J}^{\D_{i}}G_{d-\D,J}^{\D_{i}}(u,v)\label{eq:4}
\end{align}
which is invariant under shadow transformation $\D\ra d-\D$. Since $\mu$ is integrated over $\R$ in \eqref{eq:1}, we can just take one component of \eqref{eq:4}, absorb the coefficient $\g_{\D,J}^{\D_{i}}$ into $a(\D,J)$, and symmetrize it as $a(\D,J)=a(d-\D,J)$.  In $u\ra0$ limit, $G_{\D,J}^{\D_{i}}(u,v)$  has a simple hypergeometric form
\begin{equation}
G_{\D,J}^{\D_{i}}(u,v)\ra u^{(\D-J)/2}(1-v)^{J}{}_{2}F_{1}(\f{\D+J-\D_{12}}2,\f{\D+J+\D_{34}}2;\D+J;1-v)\label{eq:6}
\end{equation}
where $\D_{ij}\equiv \D_i-\D_j$. 

For the two-point function of composite operator \eqref{eq:2.3}, we will take $\phi_1=\phi_2=W$ and $\phi_3=\phi_4=V$ in \eqref{eq:1} with coordinate \eqref{eq5}. To further simplify the expression, let us consider the smearing function
$f_{W,V}$ defining $\mO_{W,V}$ by (\ref{eq:436-1}) only depends
on Rindler time distance $T_{12}$ and $T_{34}$. Precisely, in Rindler coordinate we set 
$\r_{12}=\r_{34}=\vec{x}_{\perp,12}=\vec{x}_{\perp,34}=0$ and $f_{W,V}(T_{12,34})$ only depends on $T_{12},T_{34}>0$. Expanding the cross ratio in leading order of $\e$ leads to 
\begin{align}
u & \app\f{4r^{2}}{(\cosh(\r-\r')+\f 12x_{\perp}^{2}e^{-\r-\r'}-\cosh T_{\pm})^{2}}\label{eq:484-1}\\
1-v & \app4r\f{(\cosh(\r-\r')+\f 12x_{\perp}^{2}e^{-\r-\r'})\cosh T_{\pm}-1}{(\cosh(\r-\r')+\f 12x_{\perp}^{2}e^{-\r-\r'}-\cosh T_{\pm})^{2}}\label{eq:480-2}
\end{align}
where $r\equiv T_{12}T_{34}/4\sim O(\e^{2})>0$ and $T_{\pm}\equiv T\pm\i\d$ with real $\d\in[0,\pi]$. Since $u$ is small, the conformal block \eqref{eq:6} is valid for our consideration. To justify this approximation for all Rindler locations, we should
require $\e/\d\ll1$, which physically means that we can regard $\mO_{W,V}$
as a very local composite operator. In the momentum space dual to $(T,\r,\vec x_{\perp})$
variables, this approximation should only change the UV structure
but not the IR regime that we are mostly interested in.

\begin{figure}
\begin{centering}
\includegraphics[width=8cm]{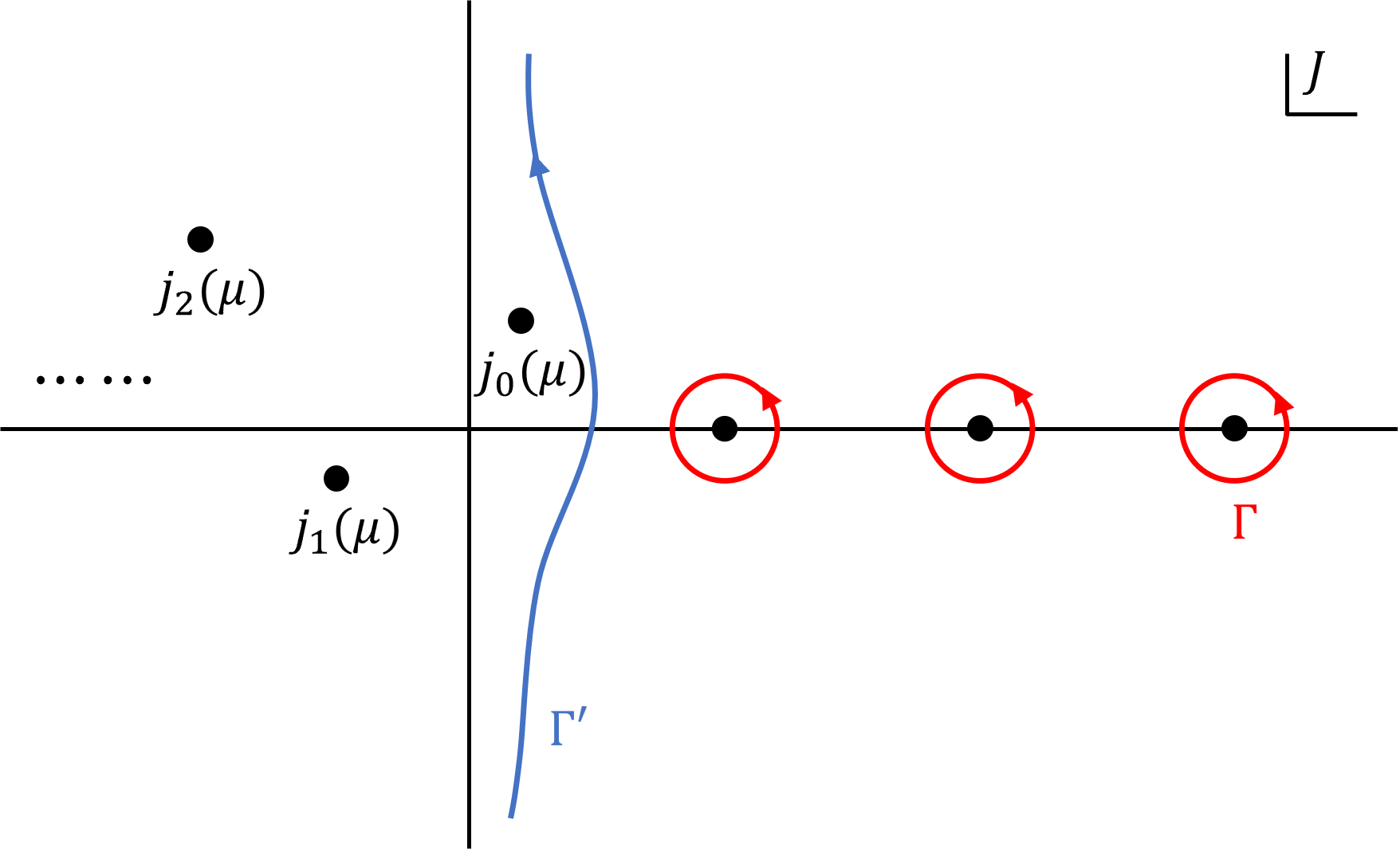}
\par\end{centering}
\caption{The contour change of $J$ integral from $\G$ (spin $J$ integers) to $\G'$ enclosing Regge trajectories $j_0(\mu),j_1(\mu),\cdots $ in the Sommerfeld-Watson resummation.\label{fig:f1} }
\end{figure}

Summing over all spins can be formulated in terms of Sommerfeld-Watson resummation \cite{sommerfeld1949partial,watson1918diffraction}. As illustrated in FIG. \ref{fig:f1}, the infinite sum over $J$ can be rewritten as a contour integral of $J$ along $\G$ around each spin $J$ integers with a kernel $\f{1}{2(\pm e^{\pi \i J}- 1)}$ or $\f{1}{2(1\mp e^{-\pi \i J})}$, where $\mp$ sign is for even/odd spins. In conformal Regge theory, even spins and odd spins are usually organized into separate Regge trajectories and the leading Regge trajectory consists of even spins with lowest twist. Therefore, we will mainly focus on even spins though the computation can be easily generalized to odd spins. The choice of kernel depends on the $J$ dependence as we analytically continue $T$ to large Lorentzian time separation. One can easily check from \eqref{2.7} that $u$ and $1-v$ circle around the origin anticlockwise for one round if $\Im T\in[-\pi,0],\Re T\ra +\infty$ or $\Im T\in[0,\pi],\Re T\ra -\infty$. To deform the contour $\G$ to $\G'$ enclosing Regge trajectories in FIG. \ref{fig:f1}, we need to use $\f{1}{2(1- e^{-\pi \i J})}$ to make sure the infinite arc $J$ integral on the right half plane vanishes. Similarly, for   $\Im T\in[-\pi,0],\Re T\ra -\infty$ or $\Im T\in[0,\pi],\Re T\ra +\infty$, we need to use the other kernel $\f{1}{2(e^{\pi \i J}- 1)}$. It turns out that the result after resummation is
\begin{equation}
\avg{\mO_{W}(x)\mO_{V}(x')}=\int_{B_{\e}}f_{W}f_{V}g_{W}g_{V}\int_{-\infty}^{\infty}\f{d\mu}{2\pi}\ointctrclockwise_{\G'}\f{dJa(\mu^{2},J)e^{-\i\f{\pi}2J\sgn(\Re T)\sgn(\Im T)}}{4(-\i)\sin(\pi J/2)}G_{\D,J}^{\D_{i}}(u,v) \label{eq:480-1}
\end{equation}

Since \eqref{eq:480-1} is essentially a time-ordered four-point function, the hypergeometric function does not go to the second plane through the branch cut from $v=-\infty$ to $v=0$, and we can expand the hypergeometric function in powers of $1-v$.
Taking (\ref{eq:484-1}) and (\ref{eq:480-2}), we have 
\begin{equation}
G_{\D,J}^{\D_{i}}(u,v)=\sum_{n=0}^{\infty}\f{(\f{\D+J}2)_{n}(\f{\D+J}2)_{n}}{(\D+J)_{n}n!}\f{((\cosh(\r-\r')+\f 12x_{\perp}^{2}e^{-\r-\r'})\cosh T_{\pm}-1)^{J+n}}{(\cosh(\r-\r')+\f 12x_{\perp}^{2}e^{-\r-\r'}-\cosh T_{\pm})^{\D+J+2n}}2^{\D+J+2n}r^{\D+n}\label{eq:481-1}
\end{equation}
For $n=0$, it is nothing but the two-point function \eqref{eq:484-2} of two spin-$J$
and dimension-$\D$ operators with pure timelike polarization written
in Rindler coordinate. The higher $n$ terms are the $n$-th order
descendants. Since $r\sim O(\e^{2})$ is small, we can ignore the
contribution from descendants and only focus on the $n=0$ term. Similar to \eqref{eq9}, the momentum space retarded function of the composite operator is
\begin{align}
G_{\mO_W\mO_V}^{R}(\w,k)= & \int_{-\infty}^{\infty}\f{d\mu}{2\pi}\ointctrclockwise_{\G_{+}}\f{dJa_{+}(\mu^{2},J)C_{\D,J}}{4\sin(\pi J/2)}\int_{T>|x|}dxdTe^{\i\w T-\i kx}(e^{\i\f{\pi J}2}H_{\D,J}(T_{-},x)-e^{-\i\f{\pi J}2}H_{\D,J}(T_{+},x))\label{eq:496-1}\\
C_{\D,J}\equiv & -\f{(2\pi)^{\f{d-2}2}\G(\D+1-\f d2)}{2^{1-\D-J}}\int_{B_{\e}}f_{W}f_{V}g_{W}g_{V}r^{\D}
\end{align}

After Sommerfeld-Waston resummation, $J$ is a complex number and the hypergeometric function is no longer a polynomial, which invalidates the method in Sec. \ref{app:ps-J}. Instead, 
let us first rewrite the hypergeometric function in terms of a Barnes
integral, which leads to
\begin{equation}
H_{\D,J}=\f{e^{\mp\i\pi(\i(t+\mu)+1)}}{2\pi\G(-J)\G(\f d2-1)}\int_{-\infty}^{+\infty}dt\f{\G(-J+\i t)\G(\f d2-1+J-\i t)\G(-\i t)(\cosh x\cosh T-1)^{\i t}}{\G(\f d2+\i\mu+J-\i t)(\cosh T-\cosh x)^{\i\mu+J+1}(\sinh^{2}T)^{\i t-J}}
\end{equation}
where the $t$ integral separates the poles at $\i t=J-\N$ on its
left and the poles at $\i t=\N,\f d2-1+J+\N$ on its right. The minus/plus
sign is for $T_{\mp}$ respectively. Moreover, we will expand the
$(\sinh^{2}T)^{{\it J-\i t}}$ as
\begin{equation}
(\sinh^{2}T)^{J-{\it \i t}}=\f{e^{2(J-\i t)T}}{2^{2(J-\i t)}}(1-e^{-2T})^{2(J-\i t)}=\sum_{n=0}^{\infty}\f{(-2J+2\i t)_{n}}{2^{2(J-\i t)}n!}e^{2(J-\i t-n)T}
\end{equation}
Taking these into (\ref{eq:496-1}), we can combine the phase to get
$\sin\pi(\i(t+\mu)-J/2)$ and rewrite it in terms of gamma functions,
which yields
\begin{align}
G_{\mO_W\mO_V}^{R}(\w,k) & =\int_{-\infty}^{\infty}\f{d\mu}{2\pi}\ointctrclockwise_{\G_{+}}\f{dJa_{+}(\mu^{2},J)C_{\D,J}\G(J+1)\cos(\pi J/2)}{\i\G(\f d2-1)}g(\w,k)\label{eq:499-2}
\end{align}
where we define
\begin{align}
g(\w,k)\equiv & \sum_{n=0}^{\infty}\int\f{dt}{2\pi}\f{\G(-J+\i t,\f d2-1+J-\i t,-\i t,-2J+2\i t+n)}{2^{2(J-\i t)}n!\G(\f d2+\i\mu+J-\i t,-2J+2\i t,\i(t+\mu)-\f J2,\f{2+J}2-\i(t+\mu))}\nonumber \\
 & \times\int_{T>|x|}dxdT\f{e^{\i(\w-2(t+\i J-\i n))T-\i kx}(\cosh x\cosh T-1)^{\i t}}{(\cosh T-\cosh x)^{\i\mu+J+1}}
\end{align}
where $\G(a,b,\cdots)\equiv\G(a)\G(b)\cdots$ is a short notation
of product of gamma functions.

\begin{figure}
\begin{centering}
\includegraphics[totalheight=4.5cm]{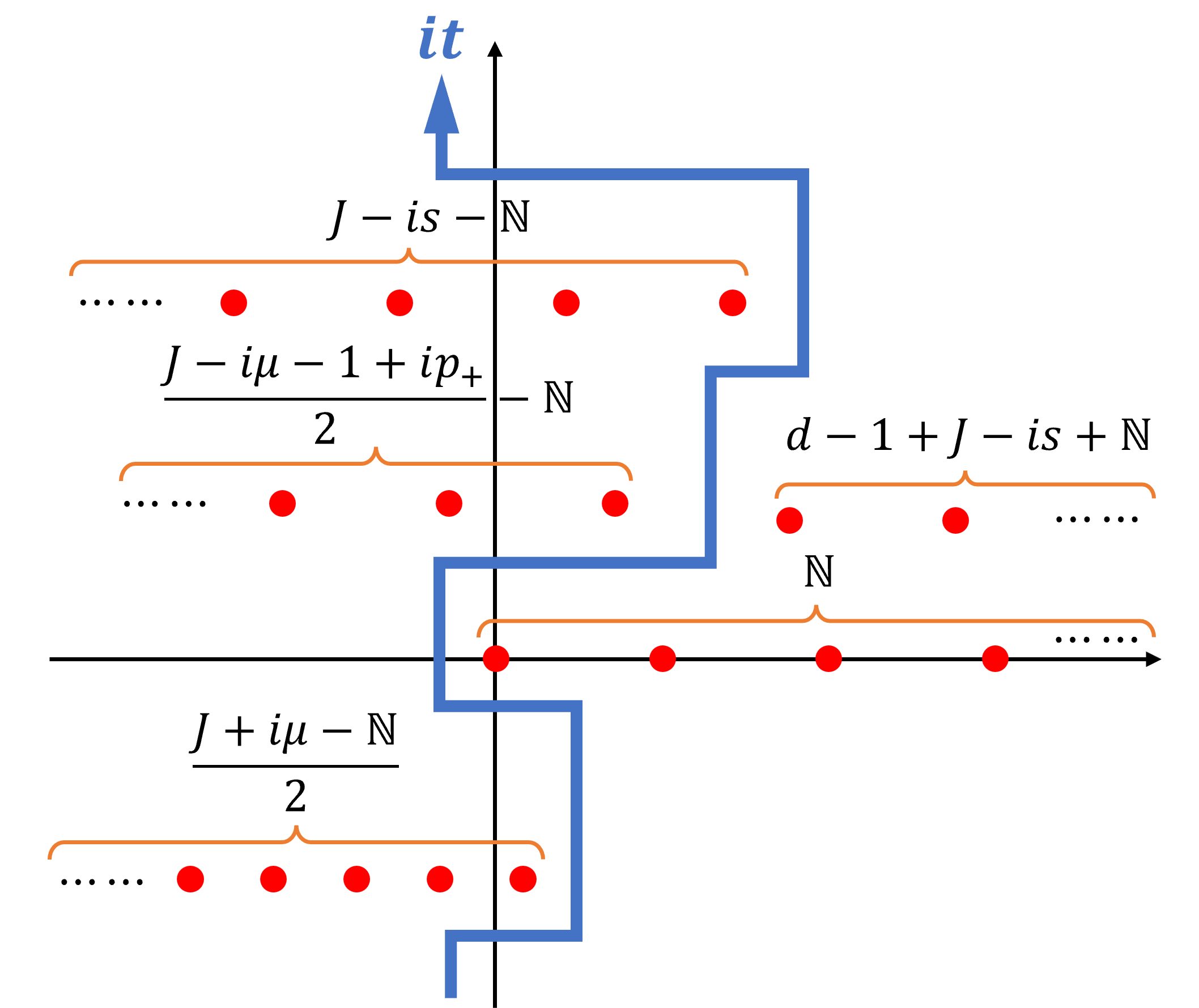}
\quad
\includegraphics[totalheight=4.5cm]{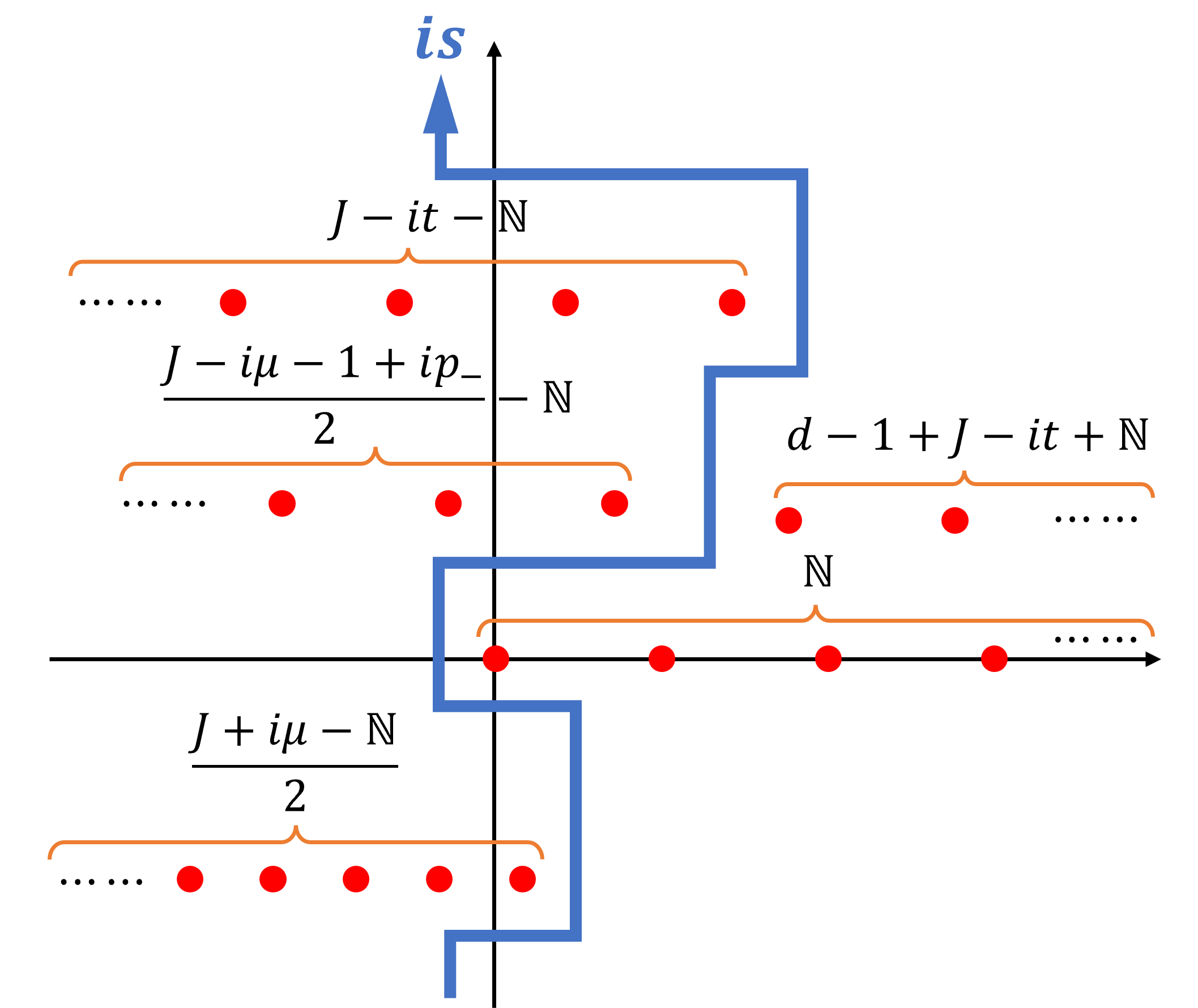}
\par\end{centering}
\caption{The blue curve is the contour of $\protect\i t$ (left) and $\protect\i s$
(right) in (\ref{eq:508-2}). Red dots are poles of the gamma function.\protect\label{fig:The-blue-curve}}
\end{figure}

To compute $g(\w,k)$, let us first shift $\w\ra\w+2(t+\i J-\i n)$
and introduce the lightcone coordinates and the conjugate momenta \eqref{s14} to rewrite
\begin{equation}
e^{\i\w T-\i kx}\f{(\cosh x\cosh T-1)^{\i t}}{(\cosh T-\cosh x)^{\i\mu+J+1}}=e^{\i p_{-}x_{+}+\i p_{+}x_{-}}\f{(\sinh^{2}x_{+}+\sinh^{2}x_{-})^{\i t}}{(2\sinh x_{+}\sinh x_{-})^{\i\mu+J+1}}
\end{equation}
Using the integral representation
\begin{equation}
(\sinh^{2}x_{+}+\sinh^{2}x_{-})^{\i t}=\int_{-\infty}^{\infty}ds\f{\G(-\i t+\i s)\G(-\i s)}{2\pi\G(-\i t)}(\sinh^{2}x_{+})^{\i(t-s)}(\sinh^{2}x_{-})^{\i s}
\end{equation}
where the $\i s$ contour separates the poles at $\i s=\i t-\N$ on
its left and the poles at $\i s=\N$ on its right. It follows that
\begin{align}
 & \int_{T>|x|}dxdTe^{\i\w T-\i kx}\f{(\cosh x\cosh T-1)^{\i t}}{(\cosh T-\cosh x)^{\i\mu+J+1}}\nonumber \\
= & \f 1{2^{\i\mu+J}}\int_{-\infty}^{\infty}ds\f{\G(-\i t+\i s)\G(-\i s)}{2\pi\G(-\i t)}\int_{0}^{\infty}\f{dx_{\pm}e^{\i p_{-}x_{+}+\i p_{+}x_{-}}}{(\sinh x_{+})^{\i(\mu-2t+2s)+J+1}(\sinh x_{-})^{\i(\mu-2s)+J+1}}\nonumber \\
= & \int_{-\infty}^{\infty}ds\f{\G(-\i t+\i s,-\i s,\i(2t-2s-\mu)-J,\i(2s-\mu)-J,\f{\i(\mu-2t+2s)+J+1-\i p_{-}}2,\f{\i(\mu-2s)+J+1-\i p_{+}}2)}{2^{\i(2t-\mu)-J}(2\pi)\G(-\i t,\f{\i(2t-2s-\mu)-J+1-\i p_{-}}2,\f{\i(2s-\mu)-J+1-\i p_{+}}2)}\label{eq:505}
\end{align}
The Fourier integral of $x_{\pm}$ converges for
\begin{equation}
-1-\Im p_{+}<\Re(\i(\mu-2s)+J)<0,\quad,-1-\Im p_{-}<\Re(\i(\mu-2t+2s)+J)<0\label{eq:506}
\end{equation}
and we continue (\ref{eq:505}) to other cases. Shifting back $p_{\pm}\ra p_{\pm}-2(t+\i J-\i n)$,
we have
\begin{align}
 & g(\w,k)=\int\f{dtds}{(2\pi)^{2}}\f{\G(-J+\i t,\f d2-1+J-\i t,-\i t+\i s,-\i s,\i(2t-2s-\mu)-J,\i(2s-\mu)-J)}{2^{J-\i\mu}\G(\f d2+\i\mu+J-\i t,-2J+2\i t,\i(t+\mu)-\f J2,\f{2+J}2-\i(t+\mu))}\nonumber \\
 & \times\sum_{n=0}^{\infty}\f{\G(-2J+2\i t+n,\f{\i(\mu+2s)-J+1-\i p_{-}}2+n,\f{\i(\mu+2t-2s)-J+1-\i p_{+}}2+n)}{n!\G(\f{\i(4t-2s-\mu)-3J+1-\i p_{-}}2+n,\f{\i(2t+2s-\mu)-3J+1-\i p_{+}}2+n)}\nonumber \\
 & =\int\f{dtds}{(2\pi)^{2}}\f{\G(-J+\i(t+s),\f d2-1+J-\i(t+s),-\i t,-\i s,\i(2t-\mu)-J,\i(2s-\mu)-J)}{2^{J-\i\mu}\G(\f d2+\i\mu+J-\i(t+s),\i(t+s+\mu)-\f J2,\f{2+J}2-\i(t+s+\mu))}\nonumber \\
 & \times\G(\f{\i\mu-J+1-\i p_{-}}2+\i s,\f{\i\mu-J+1-\i p_{+}}2+\i t) \nn\\
 &\times {}_{3}\tilde{F}_{2}\left(\begin{array}{c}
\f{\i(\mu+2s)-J+1-\i p_{-}}2,\f{\i(\mu+2t)-J+1-\i p_{+}}2,2(\i t+\i s-J)\\
\f{\i(4t+2s-\mu)-3J+1-\i p_{-}}2,\f{\i(4s+2t-\mu)-3J+1-\i p_{+}}2
\end{array};1\right)\label{eq:508-2}
\end{align}
where $_{3}\tilde{F}_{2}$ is the regularized hypergeometric function,
which is an entire function of all parameters; and in the second step
we shift $t\ra t+s$ to write the equation in a more symmetric way.
The contour of $t$ and $s$ and the relative location of poles are
illustrated in FIG. \ref{fig:The-blue-curve}.

Taking (\ref{eq:508-2}) back to (\ref{eq:499-2}), we will have four
contour integrals of $t,s,J,\mu$. Since we are looking for pole-skipping
related to a higher spin operator on a Regge trajectory, let us first localize the $J$ integral
to the Regge pole of a Regge trajectory $J=j(\mu)$. Near a Regge trajectory, the OPE coefficient $a_+$ scales as
\begin{equation}
a_{+}(\mu^{2},J)\app \f 1{2\pi \i}\f{b_{+}(\mu)}{J-j(\mu)}\label{eq:117-1}
\end{equation}
The residue of $J$ at $j(\mu)$ leads to a coefficient $b_+(\mu)$. 

For the remaining three integrals of $t,s,\mu$, since the regularized
hypergeometric function is an entire function of parameters, they
will pick three residues of the poles of the 8 gamma functions that are not in the denominator of (\ref{eq:508-2}), which are in total 56 possibilities. We will scan all of them. To
be precise, we will pick three gamma functions from (\ref{eq:508-2})
and solve $s,t,\mu$ as three functions of $p_{\pm}$ at the simple poles
at non-positive integers. Then we substitute
the solutions into the rest five gamma functions and check the locus
of the poles on the $(\w,k)$ plane. As we show in the next subsection, there are exceptional pole-skipping points related to a trajectory besides those related to each higher spin operator discussed in Sec. \ref{sec:II}.

\subsection{Preliminary exploration of all pole-skipping points} \label{app:B}

Consider $d$ as a generic non-integer real number. From the scan
of 8 gamma functions in the numerator of \eqref{eq:508-2}, the pole of $\mu$ can be categorized into the
following four cases. 
\begin{enumerate}
\item Exotic case: $\mu=\i(a_{1}j(\mu)+a_{2}d+a_{3})$.
\item Chiral $p_{\pm}$-case: $\mu=\i(a_{1}j(\mu)+a_{2}d+a_{3}+\i a_{4}p_{\pm})$.
\item Frequency case: $\mu=\i(a_{2}d+a_{3}+\i a_{4}\w)$.
\end{enumerate}
where $a_{i}$ are all rational constants. We number the 8 gamma functions
in the numerator of \eqref{eq:508-2} from 1 to 8 (i.e. 1 is for $\G(-J+\i(t+s))$
and 8 is for $\G(\f{\i\mu-J+1-\i p_{+}}2+\i t)$), and Table \ref{tab:Three-cases-of}
summarizes which three choices correspond to each case. We will explore
the corresponding pole-skipping points of these four cases respectively. The analysis of this subsection only account for the pole-skipping due to individual gamma functions. There could be more sporadic pole-skipping points due to the linear combination of different gamma functions similar to the case of a spin $J$ operator. These pole-skipping points highly depend on the OPE data and are non-universal.

\begin{table}
\begin{centering}

\begin{tabular}{|c|c|}
\hline 
\begin{cellvarwidth}[h]
\centering
Cases

(total number)
\end{cellvarwidth} & Choices of gamma functions\tabularnewline
\hline 
\begin{cellvarwidth}[h]
\centering
Exotic case \\
(16)
\end{cellvarwidth} & \begin{cellvarwidth}[h]
\centering
\{1, 3, 5\}, \{1, 3, 6\}, \{1, 4, 5\}, \{1, 4, 6\}, \{1, 5, 6\}, \{2,
3, 5\}, \{2, 3, 6\}, \{2, 4, 5\}, 

\{2, 4, 6\}, \{2, 5, 6\}, \{3, 4, 5\}, \{3, 4, 6\}, \{3, 5, 6\}, \{3,
5, 7\}, \{4, 5, 6\}, \{4, 6, 8\}
\end{cellvarwidth}\tabularnewline
\hline 
\begin{cellvarwidth}[h]
\centering
$p_{+}$-case

(12)
\end{cellvarwidth} & \begin{cellvarwidth}[h]
\centering
\{1, 3, 8\}, \{1, 4, 8\}, \{1, 5, 8\}, \{2, 3, 8\}, \{2, 4, 8\}, \{2,
5, 8\}, 

\{3, 4, 8\}, \{3, 6, 8\}, \{3, 7, 8\}, \{4, 5, 8\}, \{5, 6, 8\}, \{5,
7, 8\}
\end{cellvarwidth}\tabularnewline
\hline 
\begin{cellvarwidth}[h]
\centering
$p_{-}$-case

(12)
\end{cellvarwidth} & \begin{cellvarwidth}[h]
\centering
\{1, 3, 7\}, \{1, 4, 7\}, \{1, 6, 7\}, \{2, 3, 7\}, \{2, 4, 7\}, \{2,
6, 7\}, 

\{3, 4, 7\}, \{3, 6, 7\}, \{4, 5, 7\}, \{4, 7, 8\}, \{5, 6, 7\}, \{6,
7, 8\}
\end{cellvarwidth}\tabularnewline
\hline 
\begin{cellvarwidth}[h]
\centering
Frequency case

(2)
\end{cellvarwidth} & \{1, 7, 8\}, \{2, 7, 8\}\tabularnewline
\hline 
\end{tabular}
\par\end{centering}
\caption{Four cases of gamma function choices. \protect\label{tab:Three-cases-of}}
\end{table}

\subsubsection{Exotic case}

In the exotic case, $\mu$ is not a function of $p_{\pm}$ but fixed
by a nontrivial equation. Depending on whether $a_{1}=0$ or not,
this equation could have simple solution or nontrivial solution that
depending on the trajectory $j(\mu)$. Denote this solution as $\mu_{*}$
and we will fix a point $j(\mu_{*})$ on the trajectory accordingly.
For a generic local operator in a generic CFT, its conformal dimension
$\D-d/2=\i\mu$ does not take these special values (e.g. $j(\mu_{*})$
is not an integer in general). Therefore, theses fixed value $\mu_{*}$
do \textit{not} associated to any local operator in the CFT.

Since $\mu=\mu_{*}$ and $j(\mu_{*})$ are completely fixed, the pole
line from the poles of the remaining five gamma functions are linear
in $p_{\pm}$. Due to $k\ra-k$ reflection symmetry, let us only focus
on the pole lines as a function of $p_{-}$. For the 16 choices in
Table \ref{tab:Three-cases-of}, only one of the remaining five gamma
functions leads to a potential pole line of $p_{-}$. This is equivalent
to choosing 4 gamma functions from the 8 gamma functions of \eqref{eq:508-2} to simultaneously
solve $\{t,s,\mu,p_{-}\}$ at their poles at non-positive integers.
There are in total 15 cases because two of the 16 choices are equivalent.
We list the result in Table \ref{tab:The-pole-lines}.

\begin{table}
\centering
\begin{centering}

\begin{tabular}{|c|c|c|c|}
\hline 
\# & Choices & $\i\mu_{*}=\cdots$ & (potential) pole line $\i p_{-}=\cdots$\tabularnewline
\hline 
1 & \{1,3,5,7\} & $-j(\mu_{*})+2n_{2}+n_{3}$ & $-2n_{1}+n_{3}+2n_{4}+1$\tabularnewline
\hline 
2 & \{1,3,6,7\} & $j(\mu_{*})-2n_{1}-2n_{2}+n_{3}$ & $1+2j(\mu_{*})-4n_{1}-4n_{2}+n_{3}+2n_{4}$\tabularnewline
\hline 
3 & \{1,4,5,7\} & $j(\mu_{*})-2n_{1}-2n_{2}+n_{3}$ & $1-2n_{1}+n_{3}+2n_{4}$\tabularnewline
\hline 
4 & \{1,4,6,7\} & $-j(\mu_{*})+2n_{2}+n_{3}$ & $1-2j(\mu_{*})+4n_{2}+n_{3}+2n_{4}$\tabularnewline
\hline 
5 & \{1,5,6,7\} & $\frac{1}{2}\left(-2n_{1}+n_{2}+n_{3}\right)$ & $1-2n_{1}+n_{2}+2n_{4}$\tabularnewline
\hline 
6 & \{2,3,5,7\} & $-j(\mu_{*})+2n_{2}+n_{3}$ & $-1+d+2n_{1}+n_{3}+2n_{4}$\tabularnewline
\hline 
7 & \{2,3,6,7\} & $-2+d+j(\mu_{*})+2n_{1}-2n_{2}+n_{3}$ & $-3+2d+2j(\mu_{*})+4n_{1}-4n_{2}+n_{3}+2n_{4}$\tabularnewline
\hline 
8 & \{2,4,5,7\} & $-2+d+j(\mu_{*})+2n_{1}-2n_{2}+n_{3}$ & $-1+d+2n_{1}+n_{3}+2n_{4}$\tabularnewline
\hline 
9 & \{2,4,6,7\} & $-j(\mu_{*})+2n_{2}+n_{3}$ & $1-2j(\mu_{*})+4n_{2}+n_{3}+2n_{4}$\tabularnewline
\hline 
10 & \{2,5,6,7\} & $\frac{1}{2}\left(-2+d+2n_{1}+n_{2}+n_{3}\right)$ & $-1+d+2n_{1}+n_{2}+2n_{4}$\tabularnewline
\hline 
11 & \{3,4,5,7\} & $-j(\mu_{*})+2n_{1}+n_{3}$ & $1-2j(\mu_{*})+2n_{1}+2n_{2}+n_{3}+2n_{4}$\tabularnewline
\hline 
12 & \{3,4,6,7\} & $-j(\mu_{*})+2n_{2}+n_{3}$ & $1-2j(\mu_{*})+4n_{2}+n_{3}+2n_{4}$\tabularnewline
\hline 
13 & \{3,5,6,7\} & $-j(\mu_{*})+2n_{1}+n_{2}$ & $1-2j(\mu_{*})+4n_{1}+2n_{2}-n_{3}+2n_{4}$\tabularnewline
\hline 
14 & \{4,5,6,7\} & $-j(\mu_{*})+2n_{1}+n_{3}$ & $1-2j(\mu_{*})+4n_{1}+n_{3}+2n_{4}$\tabularnewline
\hline 
15 & \{4,6,7,8\} & $-j(\mu_{*})+2n_{1}+n_{2}$ & $1-2j(\mu_{*})+4n_{1}+n_{2}+2n_{3}$\tabularnewline
\hline 
\end{tabular}
\par\end{centering}
\begin{centering}

\begin{tabular}{|c|c|c|}
\hline 
\# & zero line $\i p_{+}=\cdots$ & pole-skipping point $(\Im\w,\Im k)_{p.s.}$\tabularnewline
\hline 
1 & pole line is canceled & N/A\tabularnewline
\hline 
2 & \begin{cellvarwidth}[h]
\centering
$1-2n_{1}+n_{3}-2n_{5}$\\
$1\leq n_{5}\leq n_{3}-\min\{n_{4},2n_{1}\}$
\end{cellvarwidth} & $\begin{pmatrix}3n_{1}+2n_{2}-n_{3}-n_{4}+n_{5}-j(\mu_{*})-1\\
j(\mu_{*})-n_{1}-2n_{2}+n_{4}+n_{5}
\end{pmatrix}$\tabularnewline
\hline 
3 & pole line is canceled & N/A\tabularnewline
\hline 
4 & \begin{cellvarwidth}[h]
\centering
$1-2n_{1}+n_{3}-2n_{5}$\\
$1\leq n_{5}\leq n_{3}-\min\{n_{4},2n_{1}\}$
\end{cellvarwidth} & $\begin{pmatrix}n_{1}-2n_{2}-n_{3}-n_{4}+n_{5}+j(\mu_{*})-1\\
-j(\mu_{*})+n_{1}+2n_{2}+n_{4}+n_{5}
\end{pmatrix}$\tabularnewline
\hline 
5 & pole line is canceled & N/A\tabularnewline
\hline 
6 & pole line is canceled & N/A\tabularnewline
\hline 
7 & \begin{cellvarwidth}[h]
\centering
$d-1+2n_{1}+n_{3}-2n_{5}$\\
$1\leq n_{5}\leq n_{3}-n_{4}$
\end{cellvarwidth} & $\begin{pmatrix}2-3n_{1}+2n_{2}-n_{3}-n_{4}+n_{5}-j(\mu_{*})-\f{3d}2\\
j(\mu_{*})+n_{1}-2n_{2}+n_{4}+n_{5}+\f d2-1
\end{pmatrix}$\tabularnewline
\hline 
8 & pole line is canceled & N/A\tabularnewline
\hline 
9 & \begin{cellvarwidth}[h]
\centering
$d-1+2n_{1}+n_{3}-2n_{5}$

$1\leq n_{5}\leq n_{3}-n_{4}$
\end{cellvarwidth} & $\begin{pmatrix}-n_{1}-2n_{2}-n_{3}-n_{4}+n_{5}+j(\mu_{*})-\f d2\\
1-j(\mu_{*})-n_{1}+2n_{2}+n_{4}+n_{5}-\f d2
\end{pmatrix}$\tabularnewline
\hline 
10 & pole line is canceled & N/A\tabularnewline
\hline 
11 & pole line is canceled & N/A\tabularnewline
\hline 
12 & \begin{cellvarwidth}[h]
\centering
$1-2j(\mu_{*})+2n_{1}+2n_{2}+n_{3}-2n_{5}$

$1\leq n_{5}\leq n_{3}-n_{4},2n_{2}+n_{3}<2n_{1}$
\end{cellvarwidth} & $\begin{pmatrix}-n_{1}-3n_{2}-n_{3}-n_{4}+n_{5}+2j(\mu_{*})-1\\
-n_{1}+n_{2}+n_{4}+n_{5}
\end{pmatrix}$\tabularnewline
\hline 
13 & pole line is canceled & N/A\tabularnewline
\hline 
14 & pole line is canceled & N/A\tabularnewline
\hline 
15 & pole line is canceled & N/A\tabularnewline
\hline 
\end{tabular}
\par\end{centering}
\caption{The pole lines, zero lines, and pole-skipping points of the exotic
case. The second column records the choices of 4 gamma functions from
the 8 gamma functions to simultaneously solve $\{t,s,\mu,p_{-}\}$
at their poles at non-positive integers $-n_{i}$ for $i=1,2,3,4$.
In this table, all $n_{i}\in\protect\N$ for $i=1,2,3,4,5$. \protect\label{tab:The-pole-lines}}
\end{table}

There are a few scenarios in this table. For many choices, the potential
pole line is completely canceled by the hypergeometric function. In
these scenarios, the hypergemetric function has a form of 
\begin{equation}
_{3}\tilde{F}_{2}\left(\begin{array}{c}
-n_{4},b_{1},b_{2}\\
-n_{i}-n_{4},c
\end{array}\right)\sim\sum_{0\leq k\leq n_{4}}\f{(-n_{4})_{k}(b_{1})_{k}(b_{2})_{k}}{\G(-n_{i}-n_{4}+k)\G(c+k)}=0\label{eq:app-1}
\end{equation}
which cancels the pole line completely. For other scenarios with pole
lines, their hypergeometric functions are all in the form of
\begin{equation}
\G(X){}_{3}\tilde{F}_{2}\left(\begin{array}{c}
-n_{4},b_{1},X\\
X-n_{3},c
\end{array}\right)\sim\sum_{0\leq k\leq n_{4}}\f{(-n_{4})_{k}(b_{1})_{k}\G(X+k)}{\G(X-n_{3}+k)\G(c+k)}
\end{equation}
where $X$ is linear in $\i p_{-}$. For this function to vanish,
we need to have 
\begin{equation}
X=n_{5}\in\Z,\quad1\leq n_{5}\leq n_{3}-n_{4}
\end{equation}
which gives the zero lines. In some cases with $b_{1}$ being a non-positive
integer, we need to replace $n_{4}\ra\min\{n_{4},-b_{1}\}$. Note
that each case in Table \ref{tab:The-pole-lines} is independent
because they either have different pole lines or different zero lines.
Given the pole lines and zero lines, we can easily solve the pole-skipping
points. Note that many different choices of $n_{i}$ may lead to the
same pole line, therefore the pole-skipping points in the last column
of Table \ref{tab:The-pole-lines} are limited to the intersection
set of all integer choices of $n_{i}$ that leads to the same pole
line. This is not specified in Table \ref{tab:The-pole-lines}, and
the distribution of these pole-skipping points will be a future work.
However, it is noteworthy that these pole-skipping points are on the
$(\Im\w,\Im k)$ plane but with $\Im\w$ not at Matsubara frequencies. 

\subsubsection{Chiral case}

Since the two chiral $p_{\pm}$-cases are related to each other by
reflection $k\ra-k$, we only need to analyze the $p_{-}$-case. As
shown in Table \ref{tab:Three-cases-of}, there are 12 choices for
the $p_{-}$-case, where in each case the remaining 5 gamma functions
are in the following two types 
\begin{equation}
\G(a_{0}+\i a_{2}p_{-}+\i a_{3}p_{+}),\G(a_{0}+a_{1}j(\mu)+\i a_{2}p_{-}+\i a_{3}p_{+})\label{eq:app-4}
\end{equation}
where $a_{i}$ are constants. The first type has no $j(\mu)$ and
the second type has a $j(\mu)$. Solving $p_{-}$ will give some pole lines identical to the exotic
case, and the additional potential pole lines are listed in Table \ref{tab:Solving--from}.
Solving $p_{+}$ from (\ref{eq:app-4}) also leads to new potential pole lines. Note that in $p_{-}$-case, $\mu=\mu(p_{-})$ is a function
of $p_{-}$. Solving $j(\mu(p_{-}))$ from the second type of gamma function in \eqref{eq:app-4} leads to new types of pole lines even if $a_{2}=a_{3}=0$. 

\begin{table}
\centering
\begin{centering}

\begin{tabular}{|c|c|c|c|}
\hline 
\# & Choices & On the pole line $\i\mu=\cdots$ & (potential) pole line $\i p_{-}=\cdots$\tabularnewline
\hline 
1 & \{1,3,7,8\} & $-1+j(\mu)+\i p_{+}-2n_{2}-2n_{4}$ & $\i p_{+}+2\left(j(\mu)-n_{1}-2n_{2}+n_{3}-n_{4}\right)$\tabularnewline
\hline 
2 & \{1,4,7,8\} & $-1-j(\mu)+\i p_{+}+2n_{1}+2n_{2}-2n_{4}$ & $\i p_{+}+2\left(-j(\mu)+n_{1}+2n_{2}+n_{3}-n_{4}\right)$\tabularnewline
\hline 
3 & \{2,3,7,8\} & $-1+j(\mu)+\i p_{+}-2n_{2}-2n_{4}$ & $-2+d+2j(\mu)+\i p_{+}+2n_{1}-4n_{2}+2n_{3}-2n_{4}$\tabularnewline
\hline 
4 & \{2,4,7,8\} & $1-d-j(\mu)+\i p_{+}-2n_{1}+2n_{2}-2n_{4}$ & $2-d-2j(\mu)+\i p_{+}-2n_{1}+4n_{2}+2n_{3}-2n_{4}$\tabularnewline
\hline 
5 & \{3,4,7,8\} & $-1+j(\mu)+\i p_{+}-2n_{1}-2n_{4}$ & $\i p_{+}-2\left(n_{1}-n_{2}-n_{3}+n_{4}\right)$\tabularnewline
\hline 
6 & \{3,6,7,8\} & $-1+j(\mu)+\i p_{+}-2n_{1}-2n_{4}$ & $-1+2j(\mu)+2\i p_{+}-4n_{1}-n_{2}+2n_{3}-4n_{4}$\tabularnewline
\hline 
7 & \{4,5,7,8\} & $\frac{1}{2}\left(-1+\i p_{+}+n_{2}-2n_{4}\right)$ & $\frac{1}{2}\left(1-2j(\mu)+\i p_{+}+4n_{1}+n_{2}+4n_{3}-2n_{4}\right)$\tabularnewline
\hline 
8 & \{5,6,7,8\} & $\frac{1}{2}\left(-1+\i p_{+}+n_{1}-2n_{4}\right)$ & $\i p_{+}+n_{1}-n_{2}+2n_{3}-2n_{4}$\tabularnewline
\hline 
\end{tabular}\\
\par\end{centering}
\begin{centering}

\begin{tabular}{|c|c|c|c|c|}
\hline 
\# & $\mathsection$ & zero line $\i p_{+}=\cdots$ & At p.s. point $\i\mu_{*}=\cdots$ & pole-skipping point $(\Im\w,\Im k)_{p.s.}$\tabularnewline
\hline 
\multirow{5}{*}{1} & $a_1$ & $-\frac{d}{2}-j(\mu)-n_{1}+2n_{2}+2n_{4}-n_{5}+1$ & $-\f d2-n_{1}-n_{5}$ & $\begin{pmatrix}d/2-1+2n_{1}-n_{3}-n_{4}+n_{5}\\
j(\mu_{*})-n_{1}-2n_{2}+n_{3}-n_{4}
\end{pmatrix}$\tabularnewline
\cline{2-5}
 & $a_2$ & $-\frac{3}{2}j(\mu)+n_{1}+2n_{2}+2n_{4}-n_{5}+1$ & $-j(\mu_{*})/2+n_{1}-n_{5}$ & $\begin{pmatrix}-1+j(\mu_{*})/2-n_{3}-n_{4}+n_{5}\\
j(\mu_{*})-n_{1}-2n_{2}+n_{3}-n_{4}
\end{pmatrix}$\tabularnewline
\cline{2-5}
 & $a_3$ & $-\frac{3}{2}j(\mu)+n_{1}+2n_{2}+2n_{4}+n_{5}+2$ & $-j(\mu_{*})/2+n_{1}+n_{5}+1$ & $\begin{pmatrix}j(\mu_{*})/2-n_{3}-n_{4}-n_{5}-2\\
j(\mu_{*})-n_{1}-2n_{2}+n_{3}-n_{4}
\end{pmatrix}$\tabularnewline
\cline{2-5}
 & $b_1$ & \begin{cellvarwidth}[h]
\centering
$-2j(\mu)+4n_{2}+2n_{4}-n_{5}+1$\\
$1\leq n_{5}\leq n_{3}-\min\{n_{3},n_{4},2n_{1}\}$
\end{cellvarwidth} & $-j(\mu_{*})+2n_{2}-n_{5}$ & $\begin{pmatrix}j(\mu_{*})+n_{1}-2n_{2}-n_{3}-n_{4}+n_{5}-1\\
j(\mu_{*})-n_{1}-2n_{2}+n_{3}-n_{4}
\end{pmatrix}$\tabularnewline
\cline{2-5}
 & $b_2$ & \begin{cellvarwidth}[h]
\centering
$-2n_{1}+2n_{4}-n_{5}+1$\\
$1\leq n_{5}\leq n_{4}-\min\{n_{3},n_{4},2n_{1}\}$
\end{cellvarwidth} & $j(\mu_{*})-2n_{1}-2n_{2}-n_{5}$ & $\begin{pmatrix}3n_{1}+2n_{2}-n_{3}-n_{4}+n_{5}-1-j(\mu_{*})\\
j(\mu_{*})-n_{1}-2n_{2}+n_{3}-n_{4}
\end{pmatrix}$\tabularnewline
\hline 
\multirow{5}{*}{2} & $a_1$ & $j(\mu)-3n_{1}-2n_{2}+2n_{4}-n_{5}+\frac{2-d}{2}$ & $-\f d2-n_{1}-n_{5}$ & $\begin{pmatrix}d/2-1+2n_{1}-n_{3}-n_{4}+n_{5}\\
n_{1}+2n_{2}+n_{3}-n_{4}-j(\mu_{*})
\end{pmatrix}$\tabularnewline
\cline{2-5}
 & $a_2$ & $j(\mu)/2-n_{1}-2n_{2}+2n_{4}-n_{5}+1$ & $-j(\mu_{*})/2+n_{1}-n_{5}$ & $\begin{pmatrix}j(\mu_{*})/2-n_{3}-n_{4}+n_{5}-1\\
n_{1}+2n_{2}+n_{3}-n_{4}-j(\mu_{*})
\end{pmatrix}$\tabularnewline
\cline{2-5}
 & $a_3$ & $j(\mu)/2-n_{1}-2n_{2}+2n_{4}+n_{5}+2$ & $-j(\mu_{*})/2+n_{1}+n_{5}+1$ & $\begin{pmatrix}j(\mu_{*})/2-n_{3}-n_{4}-n_{5}-2\\
-j(\mu_{*})+n_{1}+2n_{2}+n_{3}-n_{4}
\end{pmatrix}$\tabularnewline
\cline{2-5}
 & $b_1$ & \begin{cellvarwidth}[h]
\centering
$2j(\mu)-4n_{1}-4n_{2}+2n_{4}-n_{5}+1$\\
$1\leq n_{5}\leq n_{3}-\min\{n_{3},n_{4},2n_{1}\}$
\end{cellvarwidth} & $j(\mu_{*})-2n_{1}-2n_{2}-n_{5}$ & $\begin{pmatrix}3n_{1}+2n_{2}-n_{3}-n_{4}+n_{5}-j(\mu_{*})-1\\
n_{1}+2n_{2}+n_{3}-n_{4}-j(\mu_{*})
\end{pmatrix}$\tabularnewline
\cline{2-5}
 & $b_2$ & \begin{cellvarwidth}[h]
\centering
$-2n_{1}+2n_{4}-n_{5}+1$\\
$1\leq n_{5}\leq n_{4}-\min\{n_{3},n_{4},2n_{1}\}$
\end{cellvarwidth} & $-j(\mu_{*})+2n_{2}-n_{5}$ & $\begin{pmatrix}j(\mu_{*})+n_{1}-2n_{2}-n_{3}-n_{4}+n_{5}-1\\
n_{1}+2n_{2}+n_{3}-n_{4}-j(\mu_{*})
\end{pmatrix}$\tabularnewline
\hline 
\multirow{5}{*}{3} & $a_1$ & $-j(\mu)+n_{1}+2n_{2}+2n_{4}-n_{5}$ & $n_{1}-n_{5}-1$ & $\begin{pmatrix}1-\frac{d}{2}-2n_{1}-n_{3}-n_{4}+n_{5}\\
\frac{d}{2}+J+n_{1}-2n_{2}+n_{3}-n_{4}-1
\end{pmatrix}$\tabularnewline
\cline{2-5}
 & $a_2$ & $2-\frac{d}{2}-\f 32j(\mu)-n_{1}+2n_{2}+2n_{4}-n_{5}$ & $-\frac{d}{2}-\frac{j(\mu_{*})}{2}-n_{1}-n_{5}+1$ & $\begin{pmatrix}\frac{j(\mu_{*})}{2}-n_{3}-n_{4}+n_{5}-1\\
\frac{d}{2}+j(\mu_{*})+n_{1}-2n_{2}+n_{3}-n_{4}-1
\end{pmatrix}$\tabularnewline
\cline{2-5}
 & $a_3$ & $3-\frac{d}{2}-\f 32j(\mu)-n_{1}+2n_{2}+2n_{4}+n_{5}$ & $-\frac{d}{2}-\frac{j(\mu_{*})}{2}-n_{1}+n_{5}+2$ & $\begin{pmatrix}\frac{j(\mu_{*})}{2}-n_{3}-n_{4}-n_{5}-2\\
\frac{d}{2}+j(\mu_{*})+n_{1}-2n_{2}+n_{3}-n_{4}-1
\end{pmatrix}$\tabularnewline
\cline{2-5}
 & $b_1$ & \begin{cellvarwidth}[h]
\centering
$-2j(\mu)+4n_{2}+2n_{4}-n_{5}+1$\\
$1\leq n_{5}\leq n_{3}-\min\{n_{3},n_{4}\}$
\end{cellvarwidth} & $-j(\mu_{*})+2n_{2}-n_{5}$ & $\begin{pmatrix}j(\mu_{*})-\frac{d}{2}-n_{1}-2n_{2}-n_{3}-n_{4}+n_{5}\\
\frac{d}{2}+j(\mu_{*})+n_{1}-2n_{2}+n_{3}-n_{4}+1
\end{pmatrix}$\tabularnewline
\cline{2-5}
 & $b_2$ & \begin{cellvarwidth}[h]
\centering
$d+2n_{1}+2n_{4}-n_{5}-1$\\
$1\leq n_{5}\leq n_{4}-\min\{n_{3},n_{4}\}$
\end{cellvarwidth} & $d+j(\mu_{*})+2n_{1}-2n_{2}-n_{5}-2$ & $\begin{pmatrix}2-\frac{3d}{2}-j(\mu_{*})-3n_{1}+2n_{2}-n_{3}-n_{4}+n_{5}\\
\frac{d}{2}+j(\mu_{*})+n_{1}-2n_{2}+n_{3}-n_{4}-1
\end{pmatrix}$\tabularnewline
\hline 
\multirow{5}{*}{4} & $a_1$ & $d+j(\mu)+3n_{1}-2n_{2}+2n_{4}-n_{5}-2$ & $n_{1}-n_{5}-1$ & $\begin{pmatrix}-\frac{d}{2}-2n_{1}-n_{3}-n_{4}+n_{5}+1\\
-\frac{d}{2}-j(\mu_{*})-n_{1}+2n_{2}+n_{3}-n_{4}+1
\end{pmatrix}$\tabularnewline
\cline{2-5}
 & $a_2$ & $\frac{d}{2}+\frac{j(\mu)}{2}+n_{1}-2n_{2}+2n_{4}-n_{5}$ & $-\frac{d}{2}-\frac{j(\mu_{*})}{2}-n_{1}-n_{5}+1$ & $\begin{pmatrix}\frac{j(\mu_{*})}{2}-n_{3}-n_{4}+n_{5}-1\\
-\frac{d}{2}-j(\mu_{*})-n_{1}+2n_{2}+n_{3}-n_{4}+1
\end{pmatrix}$\tabularnewline
\cline{2-5}
 & $a_3$ & $\frac{d}{2}+\frac{j(\mu)}{2}+n_{1}-2n_{2}+2n_{4}+n_{5}+1$ & $-\frac{d}{2}-\frac{j(\mu_{*})}{2}-n_{1}+n_{5}+2$ & $\begin{pmatrix}\frac{j(\mu_{*})}{2}-n_{3}-n_{4}-n_{5}-2\\
-\frac{d}{2}-j(\mu_{*})-n_{1}+2n_{2}+n_{3}-n_{4}+1
\end{pmatrix}$\tabularnewline
\cline{2-5}
 & $b_1$ & \begin{cellvarwidth}[h]
\centering
$2d+2j(\mu)+4n_{1}-4n_{2}+2n_{4}-n_{5}-3$\\
$1\leq n_{5}\leq n_{3}-\min\{n_{3},n_{4}\}$
\end{cellvarwidth} & $d+j(\mu_{*})+2n_{1}-2n_{2}-n_{5}-2$ & $\begin{pmatrix}-\frac{3d}{2}-j(\mu_{*})-3n_{1}+2n_{2}-n_{3}-n_{4}+n_{5}+2\\
-\frac{d}{2}-j(\mu_{*})-n_{1}+2n_{2}+n_{3}-n_{4}+1
\end{pmatrix}$\tabularnewline
\cline{2-5}
 & $b_2$ & \begin{cellvarwidth}[h]
\centering
$d+2n_{1}+2n_{4}-n_{5}-1$\\
$1\leq n_{5}\leq n_{4}-\min\{n_{3},n_{4}\}$
\end{cellvarwidth} & $-j(\mu_{*})+2n_{2}-n_{5}$ & $\begin{pmatrix}-\frac{d}{2}+j(\mu_{*})-n_{1}-2n_{2}-n_{3}-n_{4}+n_{5}\\
-\frac{d}{2}-j(\mu_{*})-n_{1}+2n_{2}+n_{3}-n_{4}+1
\end{pmatrix}$\tabularnewline
\hline 
\end{tabular}
\par\end{centering}
\caption{(Part I) Solving $p_{-}$ from remaining 5 gamma functions of the
chiral $p_{-}$-case. The second column records the choices of 4 gamma
functions from the 8 gamma functions to simultaneously solve $\{t,s,\mu,p_{-}\}$
at their poles at non-positive integers $-n_{i}$ for $i=1,2,3,4$.
In this table, all $n_{i}\in\protect\N$ for $i=1,2,3,4,5$. \protect\label{tab:Solving--from}}
\end{table}

\begin{table}
\centering
\begin{centering}

\begin{tabular}{|c|c|c|c|c|}
\hline 
\# & $\mathsection$ & zero line $\i p_{+}=\cdots$ & At p.s. point $\i\mu_{*}=\cdots$ & pole-skipping point $(\Im\w,\Im k)_{p.s.}$\tabularnewline
\hline 
\multirow{5}{*}{5} & $a_1$ & $-\frac{d}{2}-2j(\mu)+3n_{1}+n_{2}+2n_{4}-n_{5}+1$ & $-\frac{d}{2}-j(\mu_{*})+n_{1}+n_{2}-n_{5}$ & $\begin{pmatrix}\frac{d}{2}+2j(\mu_{*})-2n_{1}-2n_{2}-n_{3}-n_{4}+n_{5}-1\\
-n_{1}+n_{2}+n_{3}-n_{4}
\end{pmatrix}$\tabularnewline
\cline{2-5}
 & $a_2$ & $-\frac{j(\mu)}{2}+n_{1}-n_{2}+2n_{4}-n_{5}+1$ & $\frac{j(\mu_{*})}{2}-n_{1}-n_{2}-n_{5}$ & $\begin{pmatrix}\frac{j(\mu_{*})}{2}-n_{3}-n_{4}+n_{5}-1\\
-n_{1}+n_{2}+n_{3}-n_{4}
\end{pmatrix}$\tabularnewline
\cline{2-5}
 & $a_3$ & $-\frac{j(\mu)}{2}+n_{1}-n_{2}+2n_{4}+n_{5}+2$ & $\frac{j(\mu_{*})}{2}-n_{1}-n_{2}+n_{5}+1$ & $\begin{pmatrix}\frac{j(\mu_{*})}{2}-n_{3}-n_{4}-n_{5}-2\\
-n_{1}+n_{2}+n_{3}-n_{4}
\end{pmatrix}$\tabularnewline
\cline{2-5}
 & $b_1$ & \begin{cellvarwidth}[h]
\centering
$-2j(\mu)+4n_{1}+2n_{4}-n_{5}+1$\\
$1\leq n_{5}\leq n_{3}-\min\{n_{3},n_{4}\}$ or\\
$1\leq n_{5}+2n_{2}-2n_{1}\leq n_{4}-\min\{n_{3},n_{4}\}$
\end{cellvarwidth} & $-j(\mu_{*})+2n_{1}-n_{5}$ & $\begin{pmatrix}2j(\mu_{*})-3n_{1}-n_{2}-n_{3}-n_{4}+n_{5}-1\\
-n_{1}+n_{2}+n_{3}-n_{4}
\end{pmatrix}$\tabularnewline
\cline{2-5}
 & $b_2$ & \begin{cellvarwidth}[h]
\centering
$-2j(\mu)+2n_{1}+2n_{2}+2n_{4}-n_{5}+1$\\
$1\leq n_{5}\leq n_{4}-\min\{n_{3},n_{4}\}$ or\\
$1\leq n_{5}+2n_{1}-2n_{2}\leq n_{3}-\min\{n_{3},n_{4}\}$
\end{cellvarwidth} & $-j(\mu_{*})+2n_{2}-n_{5}$ & $\begin{pmatrix}2j(\mu_{*})-n_{1}-3n_{2}-n_{3}-n_{4}+n_{5}-1\\
-n_{1}+n_{2}+n_{3}-n_{4}
\end{pmatrix}$\tabularnewline
\hline 
\multirow{1}{*}{6} &  & pole line is canceled & N/A & N/A\tabularnewline
\hline 
\multirow{1}{*}{7} &  & pole line is canceled & N/A & N/A\tabularnewline
\hline 
\multirow{1}{*}{8} &  & pole line is canceled & N/A & N/A\tabularnewline
\hline 
\end{tabular}
\par\end{centering}
\caption{Part II of Table 3. \protect\label{tab:Solving--from-1}}
\end{table}

Consider solving $p_{-}$ from (\ref{eq:app-4}). In Table \ref{tab:Solving--from},
for each pole line, there are many choices of zero lines. These zero
lines are either from the gamma functions in the denominator of \eqref{eq:508-2},
or the hypergeometric function. For the former type, we label them
as $\mathsection a_i$ in the table. Since there are three gamma functions
in the denominator of \eqref{eq:508-2}, there are three $\mathsection a_i$ $(i=1,2,3)$ zero lines
for each pole line. For the latter type, we label them as $\mathsection b_i$
in the table, in which we will encounter
\begin{equation}
\G(X)\G(Y){}_{3}\tilde{F}_{2}\left(\begin{array}{c}
-n_{3},-n_{4},-2n_{1}\\
X-n_{3},Y-n_{4}
\end{array}\right)\sim\sum_{0\leq k\leq\min\{n_{3},n_{4},2n_{1}\}}\f{\G(X)\G(Y)(-n_{4})_{k}(-n_{4})_{k}(-2n_{1})_{k}}{\G(X-n_{3}+k)\G(Y-n_{4}+k)}\label{eq:app-5}
\end{equation}
where $X$ and $Y$ are both functions of $p_{+}$. For this function
to vanish, we need to have either one of 
\begin{align}
X&=n_{5}\in\Z,\quad1\leq n_{5}\leq n_{3}-\min\{n_{3},n_{4},2n_{1}\}\label{eq:app6-0}\\
Y&=n_{5}\in\Z,\quad1\leq n_{5}\leq n_{4}-\min\{n_{3},n_{4},2n_{1}\}\label{eq:app6}
\end{align}
which gives the zero lines. These two choices leads to two $\mathsection b_i$ $(i=1,2)$
zero lines for each pole line. When $-2n_{1}$ is replaced with a
generic number, we need to replace $\min\{n_{3},n_{4},2n_{1}\}$ with
$\min\{n_{3},n_{4}\}$. In Table \ref{tab:Solving--from-1}, there
are cases (e.g. \#5$\mathsection b_1$) where $X$ and $Y$ are integers
simultaneously, and we need to take both conditions (\ref{eq:app6-0}) and (\ref{eq:app6}).
The \#6, \#7 and \#8 cases in Table \ref{tab:Solving--from-1} have
no pole line due to the same scenario of (\ref{eq:app-1}). Given
the pole lines and zeros lines, we can easily solve the pole-skipping
points and the corresponding $\mu_{*}$ value. It is noteworthy that
in many rows of Table \ref{tab:Solving--from}, the $\mu_{*}$ value
obeys the same equation as the exotic case. 

Consider solving $p_{+}$ from (\ref{eq:app-4}). If we solve $p_{+}$
by
\begin{equation}
a_{0}+\i a_{2}p_{-}+\i a_{3}p_{+}=-n_{4},\text{ or }a_{0}+a_{1}j(\mu)+\i a_{2}p_{-}+\i a_{3}p_{+}=-n_{4}
\end{equation}
it is equivalent to solving $p_{-}$ if $a_{2}\neq0$ and $a_{3}\neq0$.
These cases are discussed in Table \ref{tab:Solving--from} and \ref{tab:Solving--from-1}.
The only new pole lines come from the case with $a_{2}=0$ and $a_{3}\neq0$.
Choosing 4 from 8 gamma functions to solve $\{s,t,\mu,p_{+}\}$ simultaneously
and compare with the chiral $p_{-}$-case, there are only two choices
additional to Table \ref{tab:Solving--from} and \ref{tab:Solving--from-1}.
We list them in Table \ref{tab:Solving--from-2}. It turns out that
these two cases both have pole lines completely canceled due to the
scenario of (\ref{eq:app-1}). Therefore, it does not lead any new
pole lines or pole-skipping points. 

\begin{table}
\centering
\begin{centering}

\begin{tabular}{|c|c|c|c|c|c|}
\hline 
\# & Choices & On the pole line $\i\mu=\cdots$ & (potential) pole line $\i p_{+}=\cdots$ & zero line & p.s.\tabularnewline
\hline 
1 & \{1,6,7,8\} & $\frac{1}{2}(n_{2}-2n_{3}+\i p_{-}-1)$ & $1-2n_{1}+n_{2}+2n_{4}$ & pole line is canceled & N/A\tabularnewline
\hline 
2 & \{2,6,7,8\} & $\frac{1}{2}(n_{2}-2n_{3}+\i p_{-}-1)$ & $-1+d+2n_{1}+n_{2}+2n_{4}$ & pole line is canceled & N/A\tabularnewline
\hline 
\end{tabular}
\par\end{centering}
\caption{Solving $p_{+}$ from remaining 5 gamma functions of the chiral $p_{-}$-case.
Here we only list new potential pole lines additional to Table \ref{tab:Solving--from}
and \ref{tab:Solving--from-1}. The second column records the choices
of 4 gamma functions from the 8 gamma functions to simultaneously
solve $\{t,s,\mu,p_{+}\}$ at their poles at non-positive integers
$-n_{i}$ for $i=1,2,3,4$. \protect\label{tab:Solving--from-2}}
\end{table}

Consider solving $j(\mu(p_{-}))$ from (\ref{eq:app-4}). This case
leads to potential new pole lines to previous ones only when we have
$a_{2}=a_{3}=0$. Choosing 4 from 8 gamma functions to solve $\{s,t,\mu,j(\mu)\}$
simultaneously and compare with the chiral $p_{-}$-case, there are
only two choices additional to Table \ref{tab:Solving--from}, \ref{tab:Solving--from-1}
and \ref{tab:Solving--from-2}. We list them in Table \ref{tab:Solving--from-3}.
In these two cases, since $j(\mu(p_{-}))$ is fixed, by the Regge
trajectory we have fixed $\mu(p_{-})=\mu_{*}$ and thus fixed $p_{-}=p_{-*}$.
For the zero line, it only comes from the hypergeometric function
in the form of 
\begin{equation}
_{3}\tilde{F}_{2}\left(\begin{array}{c}
-n_{4},b,-2n_{1}\\
c,X
\end{array}\right)\sim\sum_{0\leq k\leq\min\{n_{4},2n_{1}\}}\f{(-n_{4})_{k}(b)_{k}(-2n_{1})_{k}}{\G(X+k)\G(c+k)}
\end{equation}
where $X$ is linear in $p_{+}$. For this function to vanish, we
need to have 
\begin{equation}
X=-n_{5}\in-\Z,\quad n_{5}\geq\min\{n_{4},2n_{1}\}\label{eq:6-1}
\end{equation}
which gives the zero lines. For the case with $-2n_{1}$ replaced
with a non-integer, we just need to replace $\min\{n_{4},2n_{1}\}$
with $n_{4}$. It is easy to see that \#1 in Table \ref{tab:Solving--from-3}
is exactly the case related to local higher spin operators, which
is discussed in Sec. \ref{sec:II}.

\begin{table}
\centering
\begin{centering}
\begin{tabular}{|c|c|c|c|}
\hline 
\# & Choices & On the pole line $\i\mu_{*}=\cdots$ & (potential) pole line $j(\mu_{*})=\cdots$\tabularnewline
\hline 
1 & \{1,3,4,7\} & $n_{1}+n_{2}-n_{3}-2n_{4}+\i p_{-*}-1$ & $n_{1}+n_{2}+n_{3}$\tabularnewline
\hline 
2 & \{2,3,4,7\} & $-\frac{d}{2}-n_{1}+n_{2}-n_{3}-2n_{4}+\i p_{-*}$ & $1-d/2-n_{1}+n_{2}+n_{3}$\tabularnewline
\hline 
\end{tabular}\\
\par\end{centering}
\begin{centering}

\begin{tabular}{|c|c|c|}
\hline 
\# & zero line $\i p_{+}=\cdots$ & pole-skipping point $(\Im\w,\Im k)_{p.s.}$\tabularnewline
\hline 
\multirow{1}{*}{1} & \begin{cellvarwidth}[h]
\centering
$2\left(-2n_{1}-n_{2}+n_{3}+n_{4}+n_{5}+1\right)-\i p_{-*}$\\
$n_{5}\geq\min\{n_{4},2n_{1}\}$
\end{cellvarwidth} & $\begin{pmatrix}2n_{1}+n_{2}-n_{3}-n_{4}-n_{5}-1\\
\i\mu_{*}+n_{1}+n_{4}-n_{5}
\end{pmatrix}$\tabularnewline
\hline 
\multirow{1}{*}{2} & \begin{cellvarwidth}[h]
\centering
$2\left(d+2n_{1}-n_{2}+n_{3}+n_{4}+n_{5}-1\right)-\i p_{-*}$\\
$n_{5}\geq n_{4}$
\end{cellvarwidth} & $\begin{pmatrix}-d-2n_{1}+n_{2}-n_{3}-n_{4}-n_{5}+1\\
-\frac{d}{2}+\i\mu_{*}-n_{1}+n_{4}-n_{5}+1
\end{pmatrix}$\tabularnewline
\hline 
\end{tabular}
\par\end{centering}
\caption{Solving $j(\mu(p_{-}))$ from remaining 5 gamma functions of the chiral
$p_{-}$-case. The second column records the choices of 4 gamma functions
from the 8 gamma functions to simultaneously solve $\{t,s,\mu,j(\mu)\}$
at their poles at non-positive integers $-n_{i}$ for $i=1,2,3,4$.
In this table, all $n_{i}\in\protect\N$ for $i=1,2,3,4,5$. \protect\label{tab:Solving--from-3}}
\end{table}

Similar to the exotic case, in Table \ref{tab:Solving--from}, \ref{tab:Solving--from-1},
\ref{tab:Solving--from-2} and \ref{tab:Solving--from-3}, the parameter
choice of $n_{i}$ for the pole-skipping points must be the intersection
set of $n_{i}$ that leads to the same pole line and zero line (i.e.
for each row of the tables labeled by $\#$ and $\mathsection$). This is not specified in the tables,
but we again find that all pole-skipping points of the chiral case
are on the $(\Im\w,\Im k)$ plane but with $\Im\w$ not at Matsubara frequencies, except the \#1 choice in Table
\ref{tab:Solving--from-3} (recall that $d$ is a generic non-integer
real number). 

\subsubsection{Frequency case}

For the frequency case, as $\mu$ is a function of $\w=(p_{-}+p_{+})/2$,
we can solve either $p_{-}$, $p_{+}$ or $j(\mu(\w))$ at the pole
of the remaining 5 gamma functions. The result of solving $p_{-}$
is given in Table \ref{tab:Solving--from-3-1}, where there are two
additional choices but both potential pole lines are canceled due
to (\ref{eq:app-1}). Solving $p_{+}$ or $j(\mu(\w))$ neither leads
to new choices. 

\begin{table}
\centering
\begin{centering}

\begin{tabular}{|c|c|c|c|c|c|}
\hline 
\# & Choices & On the pole line $\i\mu=\cdots$ & (potential) pole line $\i p_{-}=\cdots$ & zero line & p.s.\tabularnewline
\hline 
1 & \{1,5,7,8\} & $\frac{1}{2}\left(n_{2}-2n_{4}+\i p_{+}-1\right)$ & $-2n_{1}+n_{2}+2n_{3}+1$ & pole line is canceled & N/A\tabularnewline
\hline 
2 & \{2,5,7,8\} & $\frac{1}{2}\left(n_{2}-2n_{4}+\i p_{+}-1\right)$ & $d+2n_{1}+n_{2}+2n_{3}-1$ & pole line is canceled & N/A\tabularnewline
\hline 
\end{tabular}

\par\end{centering}
\caption{Solving $p_{-}$ from remaining 5 gamma functions of the frequency
case. The second column records the choices of 4 gamma functions from
the 8 gamma functions to simultaneously solve $\{t,s,\mu,p_{-}\}$
at their poles at non-positive integers $-n_{i}$ for $i=1,2,3,4$.
In this table, all $n_{i}\in\protect\N$ for $i=1,2,3,4,5$. \protect\label{tab:Solving--from-3-1}}
\end{table}

\subsubsection{Relative location of the exceptional pole-skipping points to a Regge trajectory} \label{app:B4}

Here we just give an example of the exceptional pole-skipping points to show that its relative location on the $(\Im \w ,\Im k)$ plane to a trajectory is model-dependent. They could be either higher or lower than the trajectory.

Take $\#1\mathsection b_1$ in Table \ref{tab:Solving--from}. On the pole line, $\mu$ is a function of $p_+$
\be 
\i\mu(p_+)=-1+j(\mu(p_+))+\i p_+-2m,\quad m\equiv n_2+n_4
\ee
and the pole line is 
\be 
\i p_-=\i p_{+}+2\left(j(\mu(p_+))-m-n\right),\quad n\equiv n_1+n_2-n_3
\ee
To fix a pole line, we need to fix both $m$ and $n$ since $j(\mu)$ is generally a nonlinear function. At pole-skipping points, we have
\begin{align} 
&\i\mu_*=-j(\mu_*)+n-N,\quad N\equiv 2n_1-n-2n_3+n_5 \label{b.13}\\
1\leq &N+n+2n_3-2n_1\leq n_{3}-\min\{n_{3},m-n-n_3+n_1,2n_{1}\} \label{b.14}
\end{align}
where for a fixed zero line $N$ in \eqref{b.13} is fixed. The pole-skipping points are at
\be 
(\Im \w,\Im k)_{p.s.}=(j(\mu_{*})-1-m+N,-\i\mu_*-m-N) \label{b.15}
\ee
which is completely fixed for fixed $m,N,n$ and for all $n_{1,3}$ obeying \eqref{b.14} and $n_3\geq -n$. Since $j(\mu)$ is even in $\mu$, $(\Im \w,\Im k)=(j(\mu_*)-1,-\i\mu_*)$ is on the Regge trajectory $j(\mu)-1$. However, the relative location of \eqref{b.15} to $(j(\mu)-1,-\i \mu)$, especially which is higher in vertical direction, depends on the trajectory itself.

\section{Hamiltonian of SYK chain} \label{app:SYK}
The Hamiltonian of SYK chain is defined as
\begin{align}
& H  =\i^{q/2}\sum_{z=0}^{M-1}\Bigg[\sum_{1\leq j_{1}<\cdots <j_{q}\leq N}J_{j_{1}\cdots j_{q}}^{(z,0)}\psi_{j_{1}}^{z}\cdots\psi_{j_{q}}^{z} \nn\\
&~~~~~~+\sum_{\substack{1\leq j_{1}<\cdots <j_{q/2}\leq N\\1\leq j_{q/2+1}<\cdots <j_{q}\leq N}}J_{j_{1}\cdots j_{q/2}j_{q/2+1}\cdots j_{q}}^{(z,1)}\psi_{j_{1}}^{z}\cdots\psi_{j_{q/2}}^{z}\psi_{j_{q/2+1}}^{z+1}\cdots\psi_{j_{q}}^{z+1}\Bigg] \label{eq:6.1H}\\
&\E[J_{j_{1}\cdots j_{q/2}j_{q/2+1}\cdots j_{q}}^{(z,0)}]=\E[J_{j_{1}\cdots j_{q/2}j_{q/2+1}\cdots j_{q}}^{(z,1)}] =0,\\
&\E[(J_{j_{1}\cdots j_{q/2}j_{q/2+1}\cdots j_{q}}^{(z,0)})^{2}]=\f{2^{q-1}q!\mJ_0^2}{q^2 N^{q-1}}, \quad \E[(J_{j_{1}\cdots j_{q/2}j_{q/2+1}\cdots j_{q}}^{(z,1)})^{2}]=\f{2^{q-1}[(q/2)!]^2 \mJ_1^2}{q^2 N^{q-1}}
\end{align}
where $q$ is an even number, and the Majorana fermions $\psi_i^z$ obey $\{\psi_i^z,\psi_j^{z'}\}=\d^{zz'}\d_{ij}$ and periodic boundary condition $\psi_i^0=\psi_i^M$. For this random model, we will consider physical quantities after ensemble average. 

We can perturbatively solve it around a spatially homogeneous equilibrium saddle at inverse temperature $\b=2\pi$
\be 
\f{1}{N}\sum_i\avg{\psi^z_i(\tau)\psi^z_i(0)}=\f 1 2 \sgn(\tau)\left(\f {\cos(\lam_0\pi/2)}{\cos \f{\lam_0}{2}(\tau-\pi)}\right)^{2/q} \label{eq:3.4-sad}
\ee
where $\lam_0\leq 1$ is defined by
\be \label{lyaS}
\lam_{0}=2\sqrt{\mJ^{2}_0+\mJ_1^2}\cos\pi\lam_{0}/2 \ .
\ee
Expanding the ensemble averaged $G-\S$ action around this saddle at quadratic order, we can compute four-point functions at $O(q^2/N)$ level in the large $q\ll N$ limit \cite{Choi:2020tdj}. The function $h(p)$ is defined in terms of spatial momentum $p$ below \eqref{eq:480-3} with parameter $\eta=\mJ_{1}^{2}/(\mJ^{2}_0+\mJ_1^2)\in[0,1]$.

\end{document}